\documentclass[10pt,journal,compsoc]{IEEEtran}
\usepackage[utf8]{inputenc}
\usepackage{amsmath}
\usepackage{amsfonts}
\usepackage{amssymb}
\usepackage{algpseudocode}
\usepackage{algorithm}
\usepackage[pdftex]{graphicx}
\usepackage{epstopdf,epsfig}
\usepackage{bbold}
\usepackage{subfigure}
\usepackage{amsthm}
\usepackage{wrapfig}
\usepackage{capt-of}
\usepackage[english]{babel}
\usepackage{url}
\usepackage{color}

\ifCLASSOPTIONcompsoc
  \usepackage[nocompress]{cite}
\else
  \usepackage{cite}
\fi

\definecolor{purple}{rgb}{0.5, 0.0, 0.5}

\newtheorem{theorem}{Theorem}[section]
\newtheorem{lemma}[theorem]{Lemma}
\newtheorem{remark}[theorem]{Remark}
\newcommand{\Prob}{\mathrm{Pr}}

\begin{document}

\title{Simplifying Wireless Social Caching}
%
%
%
%

\author{Mohammed~Karmoose,
        Martina~Cardone,
        and~Christina~Fragouli
\thanks{
The authors are with the Electrical Engineering Department, University of California, Los Angeles (UCLA), CA 90095 USA (e-mail: mkarmoose@ucla.edu, martina.cardone@ucla.edu, christina.fragouli@ucla.edu).
M. Karmoose was supported by NSF under Award \#1423271. M. Cardone was supported by NSF under Award \#1321120.
}}
\IEEEtitleabstractindextext{%
\begin{abstract}
Social groups give the opportunity for a new form of caching. In this paper, we investigate how a social group of users can jointly optimize bandwidth usage, by each caching parts of the data demand, and then opportunistically share these parts among themselves upon meeting. We formulate this problem as a Linear Program (LP) with exponential complexity.
Based on the optimal solution, we propose a simple heuristic inspired by the bipartite set-cover problem that operates in polynomial time.
Furthermore, we prove a worst case gap between the heuristic and the LP solutions.
Finally, we assess the performance of our algorithm using real-world mobility traces from the MIT Reality Mining project dataset and two mobility traces that were synthesized using the SWIM model. Our heuristic performs closely to the optimal in most cases, showing a better performance with respect to alternative solutions.

\end{abstract}

}

\maketitle

\IEEEdisplaynontitleabstractindextext

%
\IEEEpeerreviewmaketitle

\IEEEraisesectionheading{\section{Introduction}\label{sec:intro}}

\IEEEPARstart{T}{oday}, a considerable fraction of data requirements in wireless networks comes from ``social groups''.
Members of a social group share common interests/goals and exhibit frequent and regular meeting patterns. 
Situations may arise where accommodating the data requirements of a social group through the wireless network is highly costly and infeasible.
Examples of such scenarios are:
(1) a group of students  
attending an online-course 
in an economically-challenged country, where it is costly
to download the material that each student needs;
(2) a group of tourists 
interested in obtaining touristic advertising and videos in a foreign country, where it is expensive to have cellular data connection; 
(3) {in the aftermath of catastrophic emergencies}, where the infrastructured networks are 
compromised and it is infeasible to establish stable connections  
with citizens.
These examples 
highlight the critical importance of reducing the dependence on infrastructured networks.
By exploiting social interactions among {group members}, it becomes possible to distribute the downloading efforts among the members who can then exchange data through local and cost-free connections.

We consider a social group of $N$ members who all wish to acquire (within a time period of duration $t$) a set of $M$ files on their smart wireless devices. These $M$ files are stored on a server to which the $N$ users have access through a wireless communication link.
Examples of this type of scenarios include co-workers downloading files needed before a meeting, conference participants downloading presentations for next sessions, students downloading class materials and sport fans downloading videos during an event. We assume that the group members have regular meeting patterns, which are correlated with the group activity (e.g., work, sport, entertainment); we model these meeting patterns as random events. In particular, we assume that with some probability, members meet each other (one or multiple times) within the period of interest.

In this work we seek to minimize the usage of the bandwidth. 
As supported by almost all smart devices today, we assume that users can connect either directly to the server through a longhaul connection (e.g., cellular), which is expensive in bandwidth, or to each other, when in physical proximity, through a local and cost-free Device-to-Device (D2D) connection (e.g., Bluetooth). 
At the beginning of the period, each member downloads a certain amount of the files through the longhaul (bandwidth expensive) connection and locally caches this information.
When two (or more) users meet, they exchange what they
have in their caches using local (cost-free) connections. 
We consider two variations: in the {\em direct} case, users share only the data they themselves have downloaded (e.g., because of liability/authentication reasons), while in the {\em indirect} case, users share both the data they themselves have downloaded as well as the data they have collected through previous encounters with other members. 
At the end of the time period of duration $t$, if a user has not received yet all the files, she will download the missing amount of data through the longhaul connection. The fundamental question we seek to answer is the following: at the beginning of the period, how much should each user download through the longhaul connection, so that the expected total usage of bandwidth within the period is minimized?

\medskip
\noindent{\bf{Related Work.}}
Distributed and cooperative caching, as a means of improving the system performance, has received considerable attention lately as summarized next.

Work in the  literature has considered the ultimate information-theoretic performance \cite{maddah2014fundamental, hachem2014multi, ji2016wireless}.
The common objective of these works is to find the optimal caching policy in a scenario where different users have different demands, where the demands may be uniform \cite{maddah2014fundamental} or not  \cite{hachem2014multi, ji2016wireless}. In all these works the amount of caching is known 
and the randomness lies in the users demands, while  in our scenario the randomness lies in the member encounters.

In a situation where a group of smartphone users, with a common and simultaneous demand, are within proximity, cooperative caching is closely related to cooperative downloading \cite{ananthanarayanan2007combine,keller2012microcast,do2011massive}. The key-ingredient of these works, similar to ours, is that each user downloads parts of the content from the server (through a longhaul connection) and then disseminates (through a Wi-Fi connection) these parts to 
users in proximity. 
Distinct from these works, 
we do not a priori assume that users within the same group will {meet} and be able to exchange data within the prescribed period.

In a scenario where cooperative caching is allowed, a natural question arises on how to
create proper incentives for
the different users to cache previously  downloaded content, which potentially is not any more useful. 
This problem has been {analyzed, e.g., in} \cite{chun2004selfish,goemans2006market,taghizadeh2013distributed}. 
In our framework, since users have a common demand, there is no rebate cost on communication within a group and {members} are always enticed to cache content, leading to  distinct algorithms. 

Cooperative caching has also been analyzed in the context of delay tolerant networks. {In \cite{reich2009age,ioannidis2010distributed},} the authors derive the optimal caching policy that maximizes the {\it{social welfare}}, i.e., the average system utility. This metric is a function of {other} factors, e.g., {users} impatience and the popularity of the files. 
{In \cite{tajbakhsh2015delay}, the authors aim to minimize the average delay and/or the average consumed energy. 
This is achieved by letting the server send random linear combinations of data packets to the users, and then - {through} heuristic algorithms - determine a set of {\it qualified} users to broadcast the transmissions to others.
}
The differentiating feature of our work, however, lies in the objective function we optimize for: 
the number of downloads from the server.
This implies that in our scenario, even if the members always have access to the longhaul link,
they would anyway wait until the end of the time period before downloading from the server. In contrast, the incentive in \cite{reich2009age,ioannidis2010distributed} would cause the users to download from the server whenever they have access, while the objective in \cite{tajbakhsh2015delay} is to minimize the average consumed energy and the average delay.

Our work is similar to data offloading in cellular delay-tolerant networks: here, the  
goal is to reduce cellular data traffic by opportunistically sharing data
among 
end-users through Terminal-To-Terminal (T2T) communications (we refer to \cite{rebecchi2015data} for a 
comprehensive 
study
on this topic). 
A widely used 
approach 
is the so-called ``subset selection'', where the central coordinator (i.e., the server) selects a subset of users 
to route the required data
to other users in the network. 
In \cite{han2012mobile}, the authors propose a target-set approach, where 
the server selects $k$ users,
with the goal to maximize the number of reached users (through T2T connections).
Since this problem is 
NP-hard, the authors propose a sub-optimal greedy heuristic. 
The authors in \cite{barbera2014data} study the regular interaction patterns 
among users to predict the VIP users (i.e., those who experience the highest number of meetings); 
these are then selected to be the local data forwarders. 
Distinct from these works: 
(i) we show that, by allowing users to cache network-coded parts of the data, the problem can be formulated as an easy-to-handle Linear Program (LP);
(ii) thanks to the rigorous mathematical formulation,
we prove an analytical performance guarantee of the proposed caching strategies;
(iii) by means of numerical evaluations on real data, we present scenarios in which our approach achieves a better performance with respect to \cite{han2012mobile}.

\medskip
\noindent {\bf{Contributions.}}
We first formulate our problem as an LP,
which allocates
amounts of data to download to each member so as to minimize the expected total cost (total number of downloads).
{Towards this goal,}
we assume that the data is coded (as in network coding \cite{Medard2006}). {
Since each user caches randomly coded data segments, it is unlikely that two different caches have the same content. 
Thus, a user receives novel information whenever she accesses a cache to which she has not had access before.
}
With this, for $N$ members, we have $2^{{N}\choose{2}}$ possible meeting patterns, each occurring with a certain probability. 
The LP {is hence of} exponential size.
We perform several simplification steps and 
prove that, in the symmetric case, i.e., when all pairs of members meet with equal probability,
the complexity of the solution 
reduces 
{to linear in $N$.}
Moreover, through an artifact, we show how the indirect case can be studied within the framework of the direct case without the need to develop a separate one.

We then show a surprising connection between our problem and the well-known {\em set-cover} problem. 
In particular, we prove that the solution of the optimal LP {is} lower bounded by the weighted sum of the solutions of several set-cover problems.
Each problem is {described} by an adjacency matrix, which is related to a possible meeting pattern among the users; the weight 
depends on
the probability that this particular meeting pattern occurs.

Next, inspired by the  structure of the solution of the optimal LP, we propose a simple polynomial-time approximation algorithm that we name AlgCov.
AlgCov is related to the bipartite set-cover problem, reduces to a closed form expression in the symmetric case, and achieves in our simulations a performance close to the optimal.
Moreover, by using approximation techniques and tools from LP duality, we analytically prove that AlgCov outputs a solution that is at most an additive worst-case gap apart from the optimal; the gap depends on the number of members and on the probability that the users meet.

Finally, we evaluate the performance of  AlgCov over real-world datasets.
We use data from  the MIT Reality Mining project \cite{pentland2009inferring}, 
as well as two synthesized mobility traces, generated by the SWIM model \cite{kosta2010small}: a simulation tool used to synthesize mobility traces of users based on their social interactions. These synthesized traces were created based on real mobility experiments conducted in IEEE INFOCOM 2005 \cite{hui2005pocket} and Cambridge in 2006 \cite{chaintreau2005pocket}.
We assess the performance  over the case where group members exhibit relatively symmetric meeting patterns (i.e., users have approximately the same expected number of users to meet) as well as asymmetric patterns (i.e., different users have different expected number of users to meet). 
For both configurations, AlgCov achieves a performance {close} to the optimal.
{AlgCov performance} is also compared with alternative solutions, e.g., the target-set {heuristic} in \cite{han2012mobile} and CopCash, a strategy which incorporates the concept of caching into the cooperative downloading approach proposed in \cite{keller2012microcast}.
This paper is based on the work in \cite{karmooseISIT2016}, with the following novel contributions: (i) proofs of {the} theorems in \cite{karmooseISIT2016}, (ii) Theorems \ref{theorem_1}, \ref{theorem_indirect} and \ref{theorem_lb}, (iii) connection to the set-cover problem, (iv) CopCash comparison, and (v) SWIM model experiments.

\medskip
\noindent {\bf{Paper Organization.}}
Section \ref{sec:proForm} introduces our problem.
Section \ref{sec:LPform} formulates the problem as an exponentially complex LP and shows that this complexity becomes linear in $N$ in the symmetric case.
Section \ref{sec::set_cover} shows the connection of the LP formulation to the set-cover problem.
Section \ref{sec:polytime_approx} proposes two polynomial time heuristics, based on which we design AlgCov in Section \ref{sec:appralgo}.
Section \ref{sec:appralgo} also derives an additive gap bound on AlgCov from the optimal solution.
Section \ref{sec:DataSet} evaluates the performance of AlgCov over real-world and synthesized traces;  Section \ref{sec:DataSet} also provides comparisons with alternative solutions.
Finally Section \ref{sec:concl} concludes the paper. 
Some of the proofs can be found in the Appendix.

\medskip
\noindent {\bf{Notation.}}
Lower and upper case letters indicate scalars, boldface
lower case letters denote vectors and boldface upper case letters indicate matrices;
calligraphic letters indicate sets;
$\left | \mathcal{A} \right|$ is the cardinality of $\mathcal{A}$, $\mathcal{A} \backslash \mathcal{B}$ is the set of elements that belong to $\mathcal{A}$ but not to $\mathcal{B}$ and ${\rm{Pow}}({\mathcal{A}})$ is the power set of $\mathcal{A}$;
$[n_1 : n_2]$ 
is the set of integers from $n_1$ to $n_2 \geq n_1$;
$[x]^+ := \max\{0, x\}$ for $x \in \mathbb{R}$;
$\mathbb{E} \left [ \cdot \right ]$ is the expected value;
$\mathbf{1}_j$ (respectively, $\mathbf{0}_j$), is a $j$-long column vector of all ones (respectively, zeros);
$\mathbf{A}^{T}$ is the transpose of the matrix $\mathbf{A}$;
$\mathbb{1}_{\{ P \} }$ is the indicator function, i.e., it is equal to $1$ when statement $P$ is true and $0$ otherwise.

\section{Setup}	 
\label{sec:proForm}

\noindent{\bf Goal.}
We consider a set $\mathcal{N}$ of $N$ users  $u_i \in \mathcal{N} \; \forall \; i \in [1 : N]$ who form a social group.  All users need to obtain the same set $\mathcal{M}$ 
of $| \mathcal{M} | = M$ information units (files), that are available from a server, within the same time
period of duration $t$.
Users can access the server through a direct longhaul wireless link that has a {\em  cost $c$} per downloaded information unit.
They can also  exchange  data with each other through a {\em cost free} D2D communication link, when (and if) they happen to physically encounter each other - when their devices can directly connect to each other, e.g. through Bluetooth.
Our goal is to minimize the average total downloading cost across the user group. Clearly, with no cooperation, the total cost is $NMc$.

\smallskip

\noindent{\bf Assumptions.} We  make the following assumptions.\\
$\bullet$  Complete encounter cache exchange. According to \cite{ekman2008working}, the average contact duration between two mobile devices is 250 seconds, sufficient for delivering
approximately 750 MBs 
using standard Bluetooth 3.0 technology.
Thus, we assume that encounters last long enough {to allow the users who meet to exchange their whole cache contents}. \\
$\bullet$  No memory constraints. Since the users demand the whole $\mathcal{M}$, we assume they have sufficient cost-free storage for it.  \\
$\bullet$  A-priori known Bernoulli distribution. 
We assume that the pairwise meetings between the users {(i) are Bernoulli distributed and (ii) occur with probabilities that are known a priori. 
Studies in the literature have been conducted to provide mobility models for users based on their social interactions (see \cite{karamshuk2011human} and references therein).
While such models are fit for simulation purposes,
they appear complex to study from an analytical point of view.
Thus, we make assumption (i) as a means to derive closed-form solutions and provide analytical performance guarantees; we also assess the validity of our derived solutions on synthesized mobility traces which use mobility models.}
Assumption (ii) can be attained
by exploiting the high regularity in human mobility \cite{gonzalez2008understanding,milgram1967small} to infer future meeting probabilities based on previous meeting occurrences.\\
$\bullet$ Delay-tolerant users. Even with a longhaul connection, 
users can endure a delay, at most of duration $t$, in data delivery so as to 
receive data via D2D communications.\\
$\bullet$ Network coded downloads. We assume that users download linear combinations of the information units \cite{Medard2006}.
\smallskip

\noindent{\bf Approach.}  
Our scheme consists of three phases, namely\\
{\em 1) Caching phase:}  before the period of duration $t$ starts, each user downloads a (possibly different) amount $x_i$ of the file set $\mathcal{M}$ using the longhaul connection at cost $cx_i$. 
In our LP formulations, we assume, without loss of generality, that $| \mathcal{M} | = M=1$, and thus $x_i$ is a fraction.\\
{\em 2) Sharing phase:} when two or more users meet, they opportunistically exchange the data they have in their caches.
We consider two separate cases: the {\em direct} sharing case, where users share data they themselves have downloaded from the server (e.g., because of liability/authentication reasons), and the {\em indirect} sharing case, where users also exchange data they have collected through previous encounters. \\
{\em 3) Post-sharing phase}: each user downloads the amount $y_i$ she may be missing from the server at a cost $cy_i$. 
In the LPs, since we assume $M=1$, we have that $0\leq y_i\leq 1$.

\smallskip

\noindent With this approach, what remains is to find the optimal caching strategy. 
For instance, it is not obvious whether a user, who we expect to meet many others, should download most of the file (so that she delivers this to them) or almost none (as she will receive it from them). 
Moreover, downloading too much may lead to unnecessary cost in the caching phase; 
downloading too little may lead to not enough cost-free  sharing opportunities, 
and thus unnecessary cost in the post-sharing phase.

\section{LP formulations}
\label{sec:LPform} 

We formulate an LP that takes as input the encounter probabilities of the users, and finds
$x_i, \: \forall i\!=\![1:N]$ that
minimize the average total cost during {the} caching and post-sharing phases.
We consider direct and indirect sharing.

\smallskip
\noindent{\bf Direct Sharing.}
{During encounters users can exchange} what they have personally downloaded from the server. Thus, whether users $u_i$ and $u_j$ 
meet each other multiple times or just once during the period of duration $t$, they can still only exchange the same data - multiple encounters do not help.

We model the encounters between the $N$ users as a random bipartite graph $(\mathcal{U},\mathcal{V},\mathcal{E})$,  where: (i) $\mathcal{U}$ contains a node for each of the $N$ users at the caching-phase, (ii) $\mathcal{V}$ contains a node for each of the $N$ users at the end of the period of duration $t$, and
(iii) an edge $e \in \mathcal{E}$ always exists between a node and itself and it exists between $(u_i,u_j)$, {with $i \neq j$,} with probability $p_{i,j}^{(t)}$; this edge captures if $u_i$ and $u_j$ meet each other (one or multiple times) during the {period of duration} $t$ and share their cache contents.
{There} are $K = 2^{{N}\choose{2}}$ realizations (configurations) of such a random graph, indexed as $k = [1:K]$. Each configuration has {an adjacency matrix} $\mathbf{A}^{(k)}$ and occurs with probability $p_k^{(t)}$. For brevity, in what follows we drop the superscript $(t)$.\footnote{With this formulation, we can directly calculate the {probabilities} $p_k, \forall k \in [1:K]$, if 
the pairwise encounters are independent and Bernoulli distributed with probabilities $p_{i,j}$; however, the Bernoulli assumption is not necessary, since the formulation only uses the {probabilities} $p_k, \forall k \in [1:K],$ that could be provided in different ways as well. We also remark that $p_k, \forall k \in [1:K]$, {not only depends on the duration $t$, but also on the start of the sharing period.}}

We denote with $\mathbf{x} = [x_{[1:N]}]^T$ the vector of the downloaded fractions 
and with $\mathbf{r}^{(k)} = [r_{[1:N]}^{(k)}]^T$ the vector of the received fractions after the sharing phase for the $k$-th configuration. 
With this we have $\mathbf{r}^{(k)} = \mathbf{A}^{(k)} \mathbf{x}$.
The cost of post-sharing downloading in the $k$-th configuration is $\mathbf{C}^{(k)} = \left[ \mathbf{C}^{(k)}(1), \hdots, \mathbf{C}^{(k)}(N)\right]^T$ with
\begin{equation*}
\mathbf{C}^{(k)}(i) = c \cdot \max \{ 0, 1-r_i\} \ \forall \ i \in[1:N].
\end{equation*}
With the goal to minimize the total cost (i.e., caching and post-sharing phases) incurred by all users, the optimal $\mathbf{x}$ becomes the solution of the following optimization problem
\begin{align*}
&\min\limits_{\mathbf{x}} & & \sum_{k =1}^K p_k \sum_{i=1}^N \mathbf{C}^{(k)}(i) + c \cdot \mathbf{1}_N^T\mathbf{x} \\
&\text{subject to } & & \mathbf{x} \geq \mathbf{0}_N,
\end{align*}
or equivalently
\begin{equation}
\label{opt_problem_original_no_relay}
\begin{aligned}
&\min\limits_{\mathbf{x},\mathbf{y}} & & f^{\text{Opt}}(\mathbf{x},\mathbf{y}) = \sum_{k =1}^K p_k \sum_{i=1}^N y_{i,k} + \mathbf{1}_N^T\mathbf{x} \\
&\text{subject to } & & \mathbf{x} \geq \mathbf{0}_N, \: \: \mathbf{y} \geq \mathbf{0}_{N \times K}, \\
&   & & \mathbf{1}_N - \mathbf{A}^{(k)} \mathbf{x} \leq \mathbf{y}_k, \forall \ k\in[1:K],
\end{aligned}
\end{equation}
where the variable ${y}_{i,k}$ represents the fraction to be downloaded in the post-sharing phase by user $i \in [1:N]$ in configuration $k \in[1:K]$ after receiving data from the users encountered in the sharing phase. Without loss of generality, we assumed $c = 1$.
The LP formulation in \eqref{opt_problem_original_no_relay} has complexity $\mathcal{O}(2^{N^2})$ (due to the $K = 2^{{N}\choose{2}}$ possible realizations over which we have to optimize), which prohibits  practical utilization - yet this formulation still serves to build intuitions and offers a yardstick for performance comparisons.

The following theorem provides an alternative formulation of the LP in \eqref{opt_problem_original_no_relay}, which reduces the complexity
to $\mathcal{O}(2^N)$.
Let $\mathbf{\pi}_v$ be any row vector of length $N$ with zeros and ones as entries.
By excluding the all-zero vector, there are $2^N - 1$ such vectors,
which we refer to as \textit{selection vectors}.
We let $\mathcal{T}$ be the set of all such vectors and $\mathcal{S}_v$ be the set of users corresponding to the selection vector $\pi_v$.
\begin{theorem}
\label{theorem_1}
Let $\pi_v \in \mathcal{T}, \: \forall \ v = [1:2^N -1]$. Then the LP in \eqref{opt_problem_original_no_relay} can be equivalently formulated as
\begin{equation}
\label{opt_problem_original_no_relay_simplified}
\begin{aligned}
&\min\limits_{\mathbf{x},\mathbf{y}} & & \sum_{v =1}^{2^N-1} \left( \sum_{u \in \mathcal{S}_v} \Prob(u \rightarrow \mathcal{S}_v) \right)  y_{v} + \mathbf{1}_N^T\mathbf{x} \\
&\text{subject to } & & \mathbf{x} \geq \mathbf{0}_{N}, \: \: \mathbf{y} \geq \mathbf{0}_{N \times K}, \\
 &   & &  {1} - \pi_{v} \mathbf{x} \leq {y}_v, \forall \ v\in[1:2^N - 1],
\end{aligned}
\end{equation}
where $\Prob(u \rightarrow \mathcal{S}_v)$ is the probability that user $u$ is connected to {all and only} the users in $\mathcal{S}_v$.
\end{theorem}
The main observation behind the proof of Theorem \ref{theorem_1} (see Appendix \ref{app:proofOfLPreduction}) is that the LP in \eqref{opt_problem_original_no_relay} has an inherent symmetry: 
the Left-Hand-Sides (LHS) of the constraints of the LP in \eqref{opt_problem_original_no_relay} are all repetitions of constraints of the form $1 - \pi_v \mathbf{x}$.
Thus, an optimal solution will let the right-hand-side of the constraints with the same LHS be equal. 
By appropriately grouping these constraints and variables together, we arrive at the LP in \eqref{opt_problem_original_no_relay_simplified} which has complexity of $\mathcal{O}(2^N)$.

\medskip

\subsubsection*{LP for the symmetric case}
\label{sec::symmetric}
We now assume that users meet pairwise with the same probability, i.e., 
 $p_{i,j} = p, \: \forall \left(i,j \right) \in [1:N]^2, i \neq j$ during the period of duration $t$. Thus, $p_k$ only depends on the number of encounters (as opposed to which exactly) that the configuration $k$ contains. 
Many realistic scenarios can be modeled as symmetric, (e.g., students in the same class, doctors in the same medical department in a hospital, military soldiers in the battlefield).
The next theorem (whose proof is provided in Appendix \ref{app:SimplificationSymmetric}) significantly simplifies the problem in \eqref{opt_problem_original_no_relay}.

\begin{theorem}
\label{theorem_2}
In the symmetric scenario, the LP in \eqref{opt_problem_original_no_relay_simplified} can be simplified to the following LP
\begin{equation}
\label{opt_problem_general_N_no_relay_simplified}
\begin{aligned}
&\min\limits_{x,y_i,i \in [1:N]} & & \sum\limits_{i=1}^N y_i 
{{N-1}\choose{i-1}} N p^{i-1} \left( 1-p \right)^{N-i}
+ N x \\
&{\rm{subject \ to }} & & x \geq 0, \quad  y_i \geq 0, \; \forall \; i \in [1:N], \\
&   & & 1-i x \leq y_i, \; \forall \;i \in [1:N]. \\
\end{aligned}
\end{equation}
\end{theorem}
{The LP in \eqref{opt_problem_general_N_no_relay_simplified} has linear complexity in $N$, i.e., the optimal solution is obtained in polynomial time.}
It is worth noting that the {symmetric assumption 
is made to get an analytical handle on the problem.} When we assess the performance on real datasets we {will relax} this assumption by requiring users to have an approximately equal average degree (i.e., number of encountered users), {as explained in Section \ref{sec:DataSet}.}

\smallskip
\noindent{\bf Indirect Sharing.}
\label{sec::indirect_sharing}
Enabling users to share both what they downloaded from the server as well as what they received from previous encounters, gives rise to interesting situations, since now, not only multiple encounters help, but also the order of the encounters matters. Assume for instance that, during the period of duration $t$, $u_1$ meets $u_2$, and later $u_2$ meets $u_3$. Now $u_3$ will have 'indirectly' received $x_1$ as well as $x_2$. If instead, $u_2$ meets $u_3$ before she meets $u_1$, then $u_3$ will only receive $x_2$, but $u_1$ will receive both $x_2$ and $x_3$.  Moreover, if $u_2$ again meets $u_3$ later during the period, $u_3$ can receive $x_1$ through this second encounter with $u_2$.

To model sequential encounters, we split the time period of duration $t$
into $T$ time segments, such that, during each segment, it is unlikely for more than one encounter opportunity to occur (note that one user can still meet multiple people simultaneously).
We then 'expand' over time our bipartite graph to a $(T+1)$-partite layered graph, by adding one layer for each time segment,
{where the $\ell$-th time segment corresponds to the duration between times $t_{\ell-1}$ and $t_{\ell}$, with $\ell \in [1:T]$}.
In contrast to the direct case,
at the end of the period of duration $t$, node $u_j$ is able to receive $x_i$ from node $u_i$,
if and only if there exists a path connecting $u_i$ at the first layer to $u_j$ at the last layer; 
$u_i$ and $u_j$ do not need to have directly met, provided that such a path exists.

Note that in the bipartite (direct) case, 
{the probability $p_{i,j}$ (respectively $p_{j,i}$) associated with the edge from user $i$ to user $j$ (respectively, from $j$ to $i$) indicates how often user $i$ shares her cache content with user $j$ (respectively, $j$ with $i$), with $p_{i,j} = p_{j,i}$.}
 Thus, using this time-expanded model, the indirect case can be readily transformed into an equivalent bipartite (direct) case, by replacing the probability of each two users meeting in the bipartite graph, with the probability of a path existing between these two users on the $(T+1)$-partite graph.
Let $t_0$ and $t_{T}$ be the time instants at which the $(T+1)$-partite graph begins and ends, respectively. Let ${\rm{P}}_N^{(T)}(u \rightarrow \mathcal{S}_v ; t_0)$ be the probability that, in the time interval between $t_0$ and $t_T$, a path exists between user $u$ and each of the users inside the set $\mathcal{S}_v$.
{We let $p_{i,j}^{(t_{\ell})}, \forall \ell=[0:T-1]$ be the probability that users $i$ and $j$ are connected between time instants $t_{\ell}$ and $t_{\ell +1}$.}
Given this, the next theorem derives the values of ${\rm{P}}_N^{(T)}(u \rightarrow \mathcal{S}_v ; t_0)$\footnote{The proof of Theorem \ref{theorem_3} is based on simple counting techniques. 
}.
\begin{theorem}
\label{theorem_indirect}
Assume a $(T+1)$-partite model, where $t_{\ell-1}$ and $t_{\ell}$ are respectively the starting and ending times of the $\ell$-th time segment, $\forall \ell \in [1:T]$.
Let $\mathcal{N}$ be the set of all users, and let $S_I \subseteq \mathcal{N}$ and $S_O \subseteq \mathcal{N}$ be two sets of users of sizes ${I}$ and ${O}$, respectively. 
Let $\mathcal{U}=S_{O} \backslash S_{I}$.
Denote with $S_{I} \rightarrow S_{O} $ the event of having the users in $S_{I}$ meeting exactly the users in $S_{O}$ and let ${\rm{P}}_N^{(n+1)}(S_{I} \rightarrow S_{O}; t_{\ell})$ be the probability of this event happening between time instants $t_{\ell}$ and $t_{\ell+n+1}$.
Then, for $(\ell,n) \in [0:T-1]^2, \ell + n \leq T-1$, this probability is given by
\begin{equation*}
\label{th1_pt1}
\begin{split}
&{\rm{P}}_N^{(n+1)}(S_{I} \rightarrow S_{O}; t_{\ell}) = 
\\ & \sum\limits_{u \in {\rm{Pow}}(\mathcal{U})} {\rm{P}}_N^{(1)}(S_{I} \rightarrow S_{I} + u; t_{\ell + n}) \cdot {\rm{P}}_N^{(n)}(S_{I} + u \rightarrow S_{O}; t_{\ell})  ,
\end{split}
\end{equation*}
where 
\begin{align*}
\text{P}_N^{(1)}(S_{I} \rightarrow S_{O}; t_n ) = \!\!\!\!\!\! \prod\limits_{a \in \mathcal{N} \backslash S_{O}} \prod\limits_{b \in S_{I}} \bar{p}^{(t_n)}_{a,b} \!\!\!\prod\limits_{c \in S_{O} \backslash S_{I}} \!\!\! \left( 1 \!-\!	\! \prod\limits_{d \in S_{I}} \bar{p}^{(t_n)}_{c,d}\right)
\end{align*} 
if $S_{{I}} \!\subseteq\! S_{{O}}$ and $\text{P}_N^{(1)}(S_{I} \rightarrow S_{O}; t_n ) =0$ otherwise, where $\bar{p}^{(t_n)}_{a,b} = 1 - p^{(t_n)}_{a,b}$.
\label{theorem_3}
\end{theorem}
Theorem \ref{theorem_3} can hence be utilized to cast the indirect sharing version of our problem as a direct sharing one. In particular, an LP of the form described in \eqref{opt_problem_original_no_relay} has to be solved, with the values of $p_k, \forall k \in [1:K]$ being replaced with those obtained from Theorem \ref{theorem_3}.
Note that these probabilities might not have the same symmetric structure as those of the direct sharing {model}\footnote{In the direct case, when user $i$ meets user $j$ with probability $p_{i,j}$, then user $j$ meets user $i$ with the same probability.}.
However, the problem formulation and the algorithms designed in next sections are readily suitable for the indirect sharing case where the graph model is not necessarily symmetric.
Thus, in the rest of the paper, for theoretical analysis
we only consider the direct case. 
However, in Section \ref{sec:DataSet}, we assess the performance of our algorithms for both the direct and indirect cases.

\section{Connection to Set-Cover Problem}
\label{sec::set_cover}

A Set-Cover (SC) problem is modeled as a bipartite graph $(\mathcal{S},\mathcal{V},\mathcal{E})$, with $\mathcal{V}$ being the set of {nodes} (i.e., the {\em universe}), $\mathcal{S}$ being a collection of sets whose union equals the universe and where an edge $e_{ij} \in \mathcal{E}$ exists between set $i \in \mathcal{S}$ and node $j \in \mathcal{V}$ if node $j$ belongs to set $i$. 
An integer LP formulation of the SC problem then finds the optimal selection variables $x_i \in \{0,1\}, \: \forall i \in [1:N]$ to minimize the number of selected sets in $\mathcal{S}$ {while 'covering' all node in $\mathcal{V}$}.

\begin{figure}[t]
 \centering
 \subfigure{
 \includegraphics[width=1.16in]{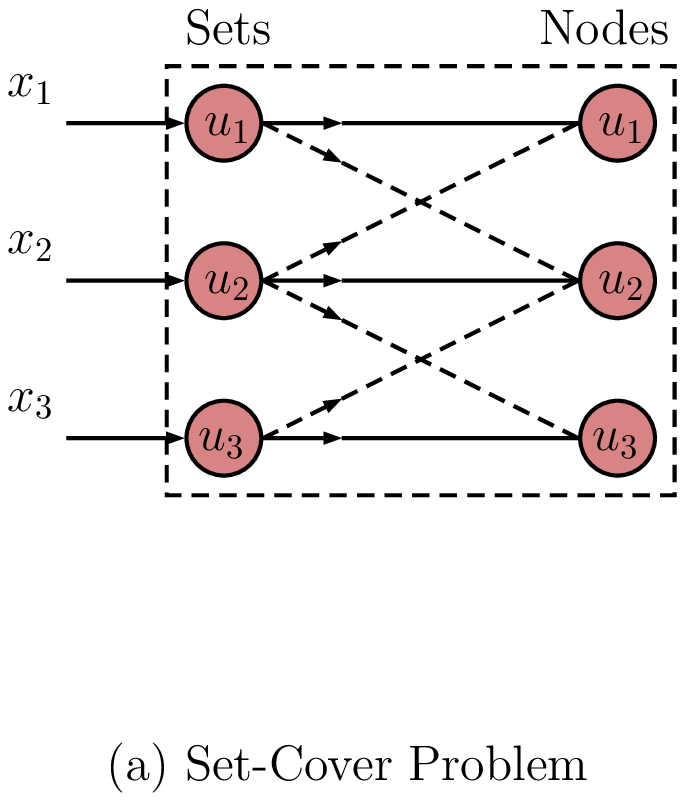}
 }
 \subfigure{
 \includegraphics[width=1.69in]{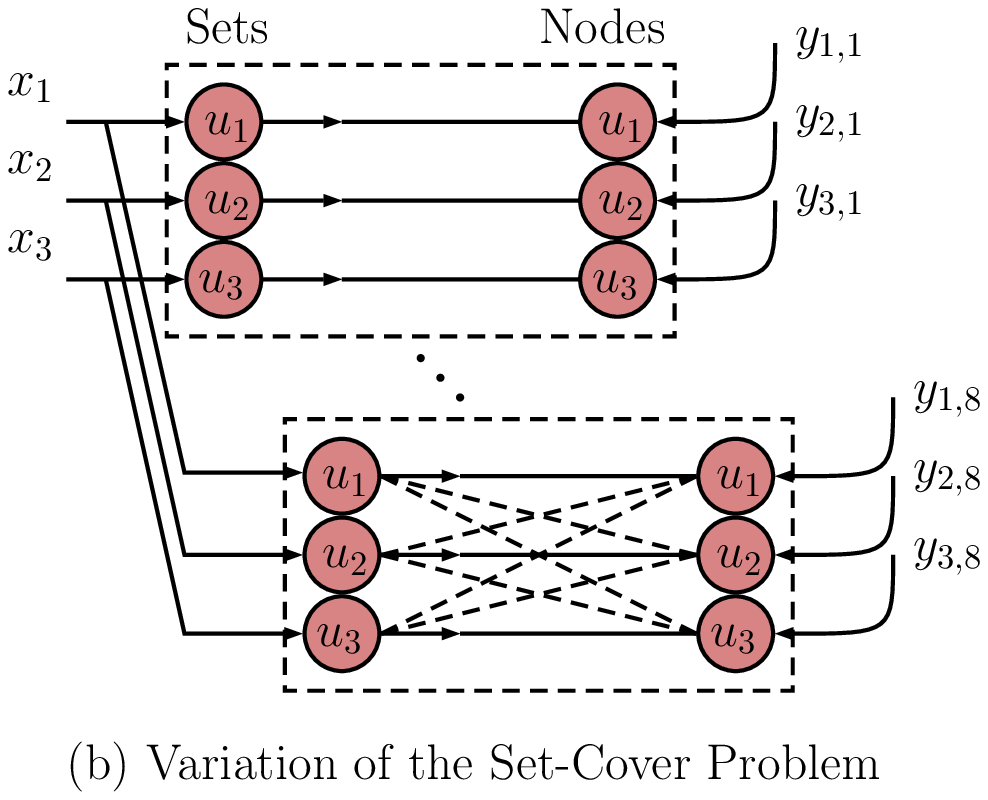}
 }
\vspace{-2mm}
   \caption{Set-cover problem and its variation.}
   \label{fig::set_cover_model}
\vspace{-4mm}
\end{figure}

One can therefore think of the LP formulation in \eqref{opt_problem_original_no_relay} as a relaxation of an integer LP, which models a variation of the SC problem.
In this variation, there are two major differences:
(i) the covering is performed on $K$ bipartite graphs, each with a different adjacency matrix $\mathbf{A}^{(k)}, k \in [1:K]$, and the same sets are selected to cover 'all' bipartite graphs;
(ii) each node can be covered by either a selected set that contains it, or an 'outside' source.
With reference to the LP in \eqref{opt_problem_original_no_relay}, the variables $x_i$ are the selection variables of the sets, and the variables $y_{i,k}$ are the outside sources of user $i$ in configuration $k$.
An illustrative example 
is given in Figure \ref{fig::set_cover_model}. A conventional SC problem is shown in Figure \ref{fig::set_cover_model}(a), where the {sets} $u_1$ and $u_3$ contain nodes ($u_1, u_2$) and ($u_2, u_3$), respectively, while set $u_2$ contains users $u_{[1:3]}$. The variables $x_{[1:3]}$ therefore determine which sets are selected for all the nodes to be covered. 
{In} this example, the set $u_2$ {covers} all the nodes.
In our variation of the SC problem in Figure \ref{fig::set_cover_model}(b), there are $8$ possible instances of bipartite graphs between $3$ sets and $3$ nodes, where the variables $x_{[1:3]}$ determine the selected sets that are used to  simultaneously cover the users in {\em{all}} {graphs}, while the variables $y_{i,k}$ are used to cover the remaining users that were not covered by the selected sets.

The following theorem proves that indeed our LP formulation in \eqref{opt_problem_original_no_relay} is closely related to the set-cover problem (see Appendix \ref{app:theorem_set_cover} for the proof).

\begin{theorem}
 \label{theorem_set_cover}
The optimal solution of the LP in \eqref{opt_problem_original_no_relay} is lower bounded by the weighted sum of the outputs of $K$ different LPs as follows.
For $k \in [1:K]$, the $k$-th LP is a relaxed SC problem over a bipartite graph with adjacency matrix $\mathbf{A}^{(k)}$. The output of the $k$-th LP is weighted by $p_k$.
\end{theorem}

\section{Polynomial Time Approximations}
\label{sec:polytime_approx}

In this section, we propose heuristics that {find} an approximate solution for the LP in \eqref{opt_problem_original_no_relay} in polynomial time.

\medskip
\noindent{\bf Inverse Average Degree (IAD).}
Consider the symmetric direct case, where users meet pairwise with the same probability $p$.
For {this} scenario, we expect that the bipartite graph has (in expectation) a constant degree of $p(N-1)+1$,
since each user, in average, meets the same number of people.
The degree, in fact, captures the number of users met in that random realization; hence, each user meets (apart from herself) the remaining $N-1$ users with equal probability $p$.

\begin{figure}[tb!]
\centering
\includegraphics[width=0.3\textwidth]{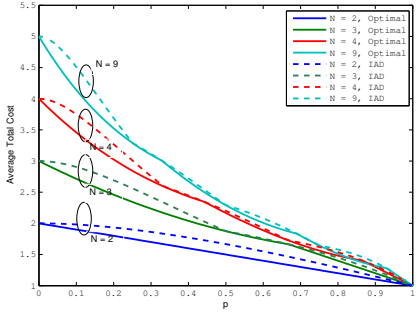}
\vspace{-2mm}
\caption{Optimal (solid lines) and IAD (dashed lines) average total cost.}
\label{fig::no_relaying_opt_vs_expected}
\vspace{-4mm}
\end{figure}

In this case, a natural heuristic {is to let each user download} $\frac{1}{\mathbb{E}(C)}$, where $C$ is a random variable corresponding to the number of people (including herself) a user meets.
Figure
\ref{fig::no_relaying_opt_vs_expected} shows 
the optimal performance (solid lines), {\em i.e.}, the solution of the LP in \eqref{opt_problem_original_no_relay}, and the performance of the caching strategy when each user downloads $\frac{1}{\mathbb{E}(C)}$ (dashed lines) for the symmetric case versus different values of $p$.
It is evident from Figure \ref{fig::no_relaying_opt_vs_expected}, that such a choice of a caching strategy closely follows the performance of the optimal solution in symmetric scenarios.
However, this approximation does not perform as well in the general (asymmetric) case.
{Consider, for example, a `star'-like configuration,} i.e., $u_1$ is highly connected to the other $N-1$ users, while the other $N-1$ users are only connected to $u_1$. 
In {this} scenario the minimum (i.e., optimal) total average cost is approximately $1$, achieved by letting $u_1$ download the whole file and then share it with the other $N-1$ users. 
In contrast, if we force $u_i \in [1:N]$ to download $\frac{1}{\mathbb{E}(C_i)}$ we would get that $u_1$ downloads $\frac{1}{N}$ (as she meets the others $N-1$ members plus herself) and $u_j, j \neq 1$ downloads $\frac{1}{2}$ (as she only meets $u_1$ plus herself). 
This would imply a total cost of $\left(\frac{1}{N} + \frac{N-1}{2} \right ) \geq 1, \forall \ N$ for the caching phase, which grows linearly with $N$ and thus can be $N$-times worse than the optimal. 
This {suggests} that the optimal search {might look like a `cover':} a set of nodes that enables to `reach' and `convey' information to all others. 
This is in line with the observations we previously made in Section \ref{sec::set_cover}.

\medskip
\noindent{\bf Probabilistic Set-Cover {(PSC)}.}
Building on this {intuition}, we propose another {heuristic} that seeks to find a form of a ``fractional covering'', where the fraction that each user downloads is a 'cover' for the users she may meet.
{In the PSC} problem \cite{beraldi2002probabilistic}, the covering constraint is replaced with a probabilistic one (i.e., the probability of covering all nodes is greater than a threshold). Here, we propose a variation of {the PSC problem} with an 'average' constraint.

We model the problem through a fully-connected bipartite graph $(\mathcal{U},\mathcal{V},\mathcal{E})$, where each edge  $u_i - v_j, \: \forall u_i \in \mathcal{U}, \; v_j \in \mathcal{V}, \ (i,j)\in [1:N]^2$ has an associated weight $p_{i,j}$, that represents how much on average $u_i$ can cover $v_j$. {We set $p_{i,i} = 1, \ \forall i\in[1:N]$, and $p_{i,j} = p_{j,i}, \ \forall (i,j) \in[1:N]^2, \ i \neq j $}. The heuristic then seeks  to associate fractional values $x_i$ to the nodes in $\mathcal{U}$ on the transmitting side, so that the sum of all $x_i$'s is minimized, while each node in $\mathcal{V}$ on the receiving side is covered, i.e., assured to receive (on average) the total amount.
This is expressed through the following LP
\begin{equation}
\label{covering_problem}
\begin{aligned}
&\min\limits_{\mathbf{x}} & & f^{\text{PSC}}(\mathbf{x}) = \sum\limits_{i=1}^N x_i \\
&\text{subject to } & &   \mathbf{P} \mathbf{x} \geq \mathbf{1}_N, \ \mathbf{x} \geq \mathbf{0}_N,
\end{aligned}
\end{equation}
where $\mathbf{P}$ is a matrix whose $(i,j)$-th entry (with $i \neq j$) is $p_{i,j}$ and with ones on the main diagonal.
This is very similar to a fractional covering problem formulation, with the only difference that $\mathbf{P}$ is not forced to be binary, but can have real components to express expectations.

The next theorem proves that, for the symmetric case, the optimal solution for the LP in \eqref{covering_problem} coincides with that of the IAD heuristic (see Appendix \ref{app:Covering} for the proof).
\begin{theorem}
\label{th:coveringSymm}
For the symmetric scenario, the optimal solution for the LP in \eqref{covering_problem}, denoted as $\mathbf{x}^{\text{PSC}}$, coincides with the IAD solution, denoted as $\mathbf{x}^{\rm{IAD}}$, {\em i.e.}, $\mathbf{x}^{\text{PSC}} = \mathbf{x}^{\rm{IAD}} = \frac{1}{\mathbb{E}(C)}\mathbf{1}_N$ where $\mathbb{E}(C) = 1+\left( N-1\right)p$.
\end{theorem}

\section{AlgCov Algorithm}
\label{sec:appralgo}

In this section we present AlgCov, a simple heuristic algorithm that combines both approaches discussed in Section~\ref{sec:polytime_approx}. 
AlgCov enables to calculate the fractions $x_i$ in polynomial time, and achieves a performance close to that of the (exponentially complex) general LP in \eqref{opt_problem_original_no_relay}.

\subsection{Motivation}
To design an algorithm that combines the merits of both heuristics presented in Section \ref{sec:polytime_approx}, one might proceed as follows: (i) compute the solution $\mathbf{x}^{\text{PSC}}$ of the {PSC} heuristic, (ii) compute the performance of this heuristic by plugging $\mathbf{x}^{\text{PSC}}$ into the LP in \eqref{opt_problem_original_no_relay} and by optimizing over $\mathbf{y}$ to find the optimal cost for this solution.
Then, repeat the same procedure for the IAD solution $\mathbf{x}^{\text{IAD}}$ and finally choose the solution with the smallest cost.

Such a solution is, in theory, {possible}. However, the process of computing the cost of each heuristic involves solving an exponentially complex LP, {prohibiting} the applicability of the heuristic. 
The following theorem helps circumvent this complexity issue (see Appendix \ref{app:theorem_lb} for the proof).

\begin{theorem}
\label{theorem_lb}
 Let $\bar{f}^{\text{Opt}}$ and $\bar{f}^{\text{PSC}}$ be the optimal values of the LPs in \eqref{opt_problem_original_no_relay} and in \eqref{covering_problem}, respectively. Then $\bar{f}^{\text{Opt}} \geq \bar{f}^{\text{PSC}}$.
\end{theorem}

Theorem \ref{theorem_lb} provides a lower bound on the optimal value of the LP in \eqref{opt_problem_original_no_relay}, and {consequently on the performance} of the solution $\mathbf{x}^{\text{PSC}}$, i.e., ${f}^{\text{Opt}} \left(\mathbf{x}^{\text{PSC}}, \mathbf{y}^{\text{PSC}}\right) \geq  \bar{f}^{\text{Opt}} \geq \bar{f}^{\text{PSC}}$, with $\mathbf{y}^{\text{PSC}}$ being obtained by evaluating the LP in \eqref{opt_problem_original_no_relay} {while setting $\mathbf{x} = \mathbf{x}^{\text{PSC}}$.}
A fairly simple lower bound on the performance of $\mathbf{x}^{\text{IAD}}$ is obtained by simply summing over the elements of the vector $\mathbf{x}^{\text{IAD}}$, i.e., ${f}^{\text{Opt}} \left(\mathbf{x}^{\text{IAD}},\mathbf{y}^{\text{IAD}} \right) \geq \mathbf{1}_N^T \mathbf{x}^{\text{IAD}}$, with $\mathbf{y}^{\text{IAD}}$ being obtained by evaluating the LP in \eqref{opt_problem_original_no_relay} {while setting $\mathbf{x} = \mathbf{x}^{\text{IAD}}$.}
As it is much simpler to compute these lower bounds, one can
{envisage to design} an algorithm which, based on the lower bounds, selects one among the two heuristics described in Section \ref{sec:polytime_approx}.

\subsection{Algorithm Description}

\begin{algorithm}[t]
 \caption{AlgCov}
 \label{fig::algcov_algorithm}
 {\fontsize{8}{10}\selectfont
 \begin{algorithmic}
  \State \textbf{Input} Pairwise probability matrix: $\mathbf{P}$
   \State \textbf{Output} AlgCov solution: $\mathbf{x}^{\text{Alg}}$
  \State Compute $\mathbf{x}^{\text{PSC}}$ - the optimal solution of the LP in \eqref{covering_problem}
   \State Compute $\mathbf{x}^{\text{IAD}}$, with $x^{\text{IAD}}_i = 1/\mathbb{E}(C_i)$, $\forall i \in [1:N]$
  \If{$\mathbf{x}^{\text{IAD}}$ is feasible in \eqref{covering_problem}}
  \ $\mathbf{x}^{\text{Alg}} = \mathbf{x}^{\text{PSC}}$
  \Else
  \ Compute $S^{\text{PSC}} = \mathbf{1}_N^T \mathbf{x}^{\text{PSC}}$ and $S^{\text{IAD}} = \mathbf{1}_N^T \mathbf{x}^{\text{IAD}}$
  \If{$S^{\text{IAD}} \leq  S^{\text{PSC}}$}
  \ $\mathbf{x}^{\text{Alg}} = \mathbf{x}^{\text{IAD}}$
  \Else \ $\mathbf{x}^{\text{Alg}} = \mathbf{x}^{\text{PSC}}$
  \EndIf
  \EndIf
 \end{algorithmic}
 }
 \end{algorithm}

AlgCov takes as input the probability matrix $\mathbf{P}$ that contains the pairwise probabilities of meeting among users, and outputs the solution $\mathbf{x}^{\text{Alg}}$ as a caching strategy. 
It first computes the two heuristic solutions, namely $\mathbf{x}^{\text{PSC}}$ and $\mathbf{x}^{\text{IAD}}$, and then, as shown in Algorithm \ref{fig::algcov_algorithm}, selects one of them as output based on $S^{\text{PSC}}$ and $S^{\text{IAD}}$, which in Theorem \ref{theorem_lb} we proved to be lower bounds on the actual performance of the heuristics.

\subsection{Analytical performance}
\noindent{\bf{Symmetric case.}} In this setting, all pairs of members meet with equal probability. 
According to Theorem \ref{th:coveringSymm}, both the IAD and the {PSC} heuristics provide the same solution, {\em i.e.}, $\mathbf{x}^{\text{Alg}} = \mathbf{x}^{\text{PSC}} = \mathbf{x}^{\text{IAD}}$.
By optimizing the objective function of the LP in \eqref{opt_problem_general_N_no_relay_simplified} over $y_i, \ i \in [1:N]$ with $\mathbf{x}=\mathbf{x}^{\rm{Alg}}$, we obtain
\begin{align}
\label{eq:ub}
f^{\rm{Opt}}(\mathbf{x}^{\text{Alg}}) &= \sum\limits_{i=1}^N \left [ 1 \!-\! \frac{i}{1\!+\!\left( N-1\right)p}\right ]^+ \cdot \nonumber \\
&{{N\!-\!1}\choose{i\!-\!1}} N p^{i-1} \left( 1-p \right)^{N\!-\!i} \!+\! \frac{N}{1\!+\!\left( N\!-\!1\right)p},
\end{align}
which is an upper bound on the optimal performance, i.e., $\bar{f}^{\rm{Opt}}\leq f^{\rm{Opt}}(\mathbf{x}^{\text{Alg}})$.
In order to provide a performance guarantee we need to understand how well $f^{\rm{Opt}}(\mathbf{x}^{\text{Alg}})$ approximates the optimal solution of the LP in \eqref{opt_problem_general_N_no_relay_simplified}.
To this end, we use the lower bound in Theorem \ref{theorem_lb}.
This lower bound, denoted as $f^{\rm{LB}}$, implies that $\bar{f}^{\rm{Opt}} \geq f^{\rm{LB}}$.
Using the structure of $\mathbf{x}^{\text{PSC}}$ for the symmetric case in Theorem \ref{th:coveringSymm}, the lower bound becomes
\begin{equation}
 \label{eq:lb}
 f^{\rm{LB}} = \frac{N}{1+\left( N-1\right)p}.
\end{equation}
By simply taking the difference between $f^{\rm{Opt}}(\mathbf{x}^{\text{Alg}})$ in \eqref{eq:ub} and $f^{\rm{LB}}$ in \eqref{eq:lb} we obtain
\begin{align*}
\mathsf{G}^{\rm{sym}} \leq \sum\limits_{i=1}^N \left [ 1 \!-\! \frac{i}{1\!+\!\left( N\!-\!1\right)p}\right ]^+ {{N\!-\!1}\choose{i\!-\!1}} N p^{i-1} \left( 1\!-\!p \right)^{N\!-\!i}.
\end{align*}
The above gap result ensures us that, in the symmetric case, the output of AlgCov is always no more than $\mathsf{G}^{\rm{sym}}$ above the optimal solution of the LP in \eqref{opt_problem_general_N_no_relay_simplified}. It is worth noting that $\mathsf{G}^{\rm{sym}}$ is only function of the number of members $N$ and of the probability $p$ that users meet.
\begin{remark}
\label{rem:WorstGap}
Through extensive numerical simulations, we observed that $\mathsf{G}^{\rm{sym}}$ is maximum for $i=1$, i.e., the probability $p^\star$ {maximizing} $\mathsf{G}^{\rm{sym}}$ is $p^\star= \frac{-N + \sqrt{5 N^2 -8N+4}}{2 \left( N-1\right)^2}$. By evaluating $\mathsf{G}^{\rm{sym}}$ in $p^\star$, we {get a worst-case (greatest) gap of} $\mathsf{G}^{\rm{sym}} \leq 0.25N$.
\end{remark}
\noindent{\bf{Asymmetric case.}} In this setting, different pairs of members meet with different probabilities. In this scenario, differently from the symmetric case analysed above, the LP in \eqref{covering_problem} does not seem to admit an easily computable closed-form solution.
For this reason, we next show how the analysis drawn for the symmetric case can be extended to find a performance guarantee for the asymmetric case as well.

In the asymmetric case, an upper bound on the solution of {AlgCov} can be found by evaluating $f^{\rm{Opt}}(\mathbf{x}^{\text{Alg}})$ in \eqref{eq:ub} in $p=p_m$, with $p_m = \min_{(i,j)\in[1:N]^2, i \neq j} \left \{p_{i,j}\right \}$. 
In other words, instead of considering different probabilities for different pairs, we set all of them to be equal to the minimum probability; this 
gives a solution which is {always worse, i.e., greater} than or equal to the optimal solution of {AlgCov} evaluated with the original (asymmetric) probability matrix. 

Similarly, a lower bound on the optimal solution of the LP in \eqref{opt_problem_original_no_relay_simplified} can be found by evaluating $f^{\rm{LB}}$ in \eqref{eq:lb} in $p=p_M$, with $p_M = \max_{(i,j)\in[1:N]^2, i \neq j} \left \{p_{i,j}\right \}$. Again, instead of considering different probabilities for different pairs, we set all of them to be equal to the maximum probability; this 
gives a solution which is {always better, i.e., smaller} than or equal to the optimal solution of the LP in \eqref{opt_problem_original_no_relay_simplified} evaluated with the original (asymmetric) probability matrix.  Thus
\begin{align*}
\mathsf{G}^{\rm{asym}} &\!\leq\! \!\sum\limits_{i=1}^N \!\!\left [ 1 \!-\! \frac{i}{1\!+\!\left( N\!-\!1\right)p_m}\right ]^+\!\! {{N\!-\!1}\choose{i\!-\!1}} N p_m^{i\!-\!1} \left( 1\!-\!p_m \right)^{N\!-\!i}
\\& \quad + \frac{N \left( N-1\right) \left( p_M - p_m\right)}{\left [ 1+\left( N-1\right) p_m\right ] \left [ 1+\left( N-1\right) p_M\right ]}.
\end{align*}
This proves that in the asymmetric case, the output of {AlgCov} is always no more than $\mathsf{G}^{\rm{asym}}$ above the optimal solution of the LP in \eqref{opt_problem_original_no_relay_simplified}. Similar to the symmetric case, also in this setting $\mathsf{G}^{\rm{asym}}$ is only a function of the number of members $N$ and of the probabilities $p_m$ and $p_M$.

\section{Data-set evaluation} \label{sec:DataSet}


In this section, we evaluate and compare the performance of our proposed solutions and algorithms using mobility traces that are obtained either from real-world experiments or via a human mobility trace synthesizer. 

\medskip
\noindent\textbf{Performance Metrics and Comparisons:} We are mainly interested in the performance of our proposed caching techniques in comparison to the conventional non-sharing solution. 
Specifically, we are interested in assessing the \textit{average total cost} (total amount  downloaded  across the caching and post-sharing phases), averaged over the experiments.  
If each user simply downloads all data, this cost is $N$. 
Versus this, we compare the performance of: \\
\noindent$\bullet$  {\em Original Formulation and AlgCov: } We calculate the average probabilities from our dataset, feed these into the LP in \eqref{opt_problem_original_no_relay} and into Algorithm \ref{fig::algcov_algorithm} that assume Bernoulli distributions, and obtain the optimal and the AlgCov heuristic solutions, respectively. 
For each experiment, we then use these caching amounts, and follow the real meeting patterns recorded in the mobility traces to exchange data and download as needed in post-sharing phase. Finally, we calculate the actual total cost, averaged over the experiments.\\
\noindent$\bullet$ {\em 1/$N$:} We evaluate the performance when each user caches $1/N$ of the data, independently of the meeting probabilities; this is a naive heuristic that does not fully exploit the opportunistic sharing possibilities.

\noindent$\bullet$ {\em CopCash:} 
We propose a modified version of the {\em cooperative sharing} algorithm originally proposed in  \cite{keller2012microcast}, where we incorporate the concept of caching.
Cooperative sharing takes advantage of the fact that nearby users, with a common demand, can collectively download the requested set of files. 
In addition, the proposed CopCash allows users to cache the received files, with the goal of exploiting next encounter opportunities to further share the data with other users.
The scheme can be described as follows:
\begin{enumerate}
\item Whenever $N$ users meet, each of them first downloads a fraction $1/N$ of the requested set of files, and then they share these parts among themselves through cost-free transmissions (e.g., Bluetooth).
\item If there exists a user (or a set of users) in the group who has already participated to a cooperative sharing instance,
she 
directly shares what she has in her cache, i.e., what she obtained from previous meetings. 
In particular, she can share only what she has downloaded (direct sharing) or the whole set of files (indirect sharing).
\item The sharing procedure continues until the end of the period of duration $t$. At this point, if a user has participated in a previous sharing instance, she will have already obtained the set of files during that sharing instance. Otherwise, she will solely download the file set.

\end{enumerate}
\begin{figure}[t]
 \centering
 \subfigure[$i=2$: Direct Sharing.]{
 \includegraphics[width=1.6in]{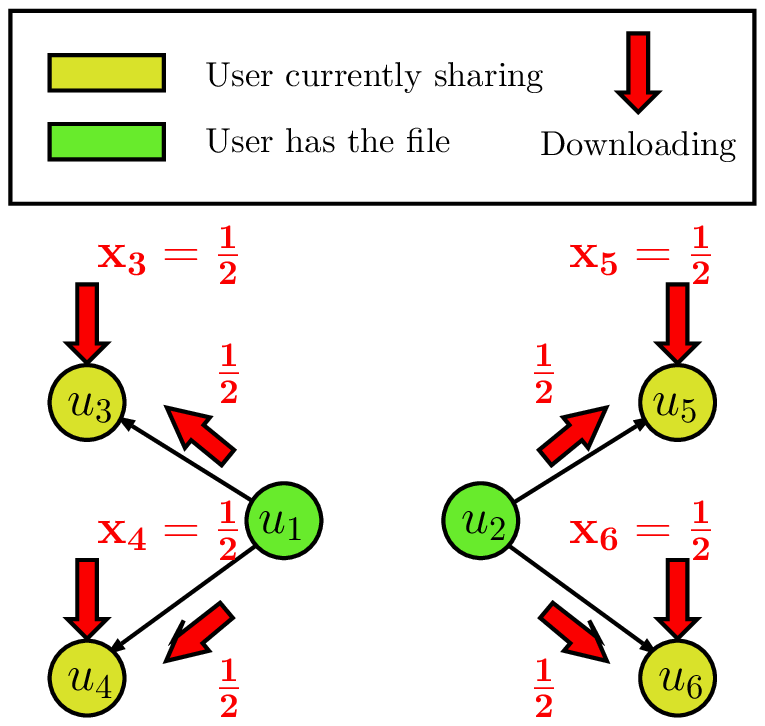}
\label{fig::CopCashb}
 }
 \subfigure[$i=2$: Indirect Sharing.]{
 \includegraphics[width=1.6in]{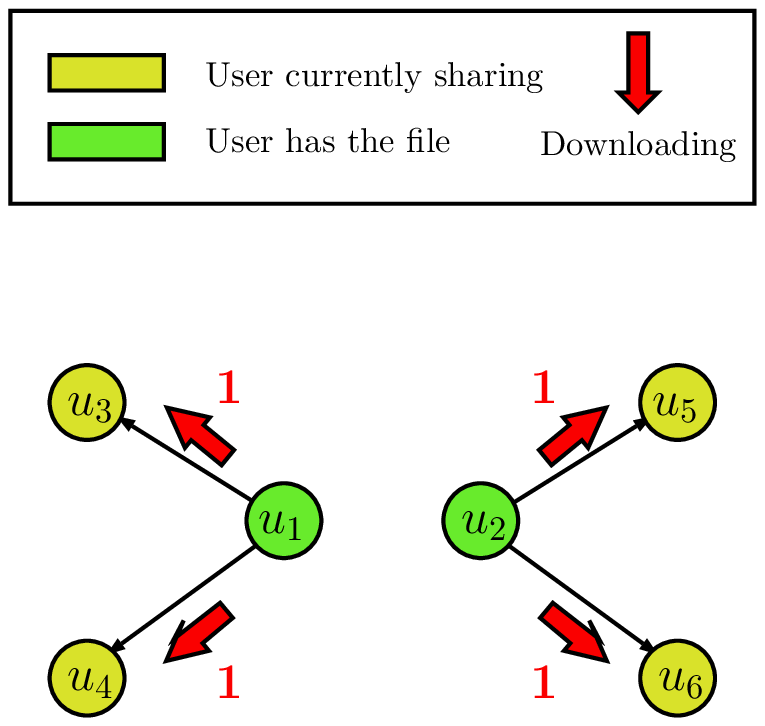}
\label{fig::CopCashc}
 }
\vspace{-1.5mm}
 \caption{An illustrative example of CopCash.}
 \label{fig::CopCash}
\vspace{-5mm}
\end{figure}

Consider the example in Figure \ref{fig::CopCash}.
Suppose that, at time instant $i=1$, $u_1$ and $u_2$ met; hence each of them downloaded $1/2$ of the file. Then, $u_1$ and $u_2$ exchanged the downloaded fractions, thus their demand was satisfied. 
At time instant $i=2$, $u_1$ meets $u_3$ and $u_4$, while $u_2$ meets $u_5$ and $u_6$. 
In the case of direct sharing - see Figure \ref{fig::CopCashb} - $u_1$ (respectively, $u_2$) shares with $u_3$ and $u_4$ (respectively, $u_5$ and $u_6$) what she has personally downloaded from the server, i.e., $1/2$ of the file; at the end of the sharing period, $u_{j}, \forall j \in [3:6]$ downloads $1/2$ of the file from the server. With this, each user has to download $1/2$ of the file.
In the case of indirect sharing - see Figure \ref{fig::CopCashc} - $u_1$ and $u_2$ share the whole set of files with the users they are connected to; in this case, $u_{j}, \forall j \in [3:6]$ does not need to download anything from the server.

\noindent$\bullet$ {\em Target-Set:} We assess the performance of the Target-Set heuristic proposed in \cite{han2012mobile} with $k=1$, i.e., the server assigns one user the task to route the data to other users. {We} only show the performance of $k=1$ since it is the case which incurs the smallest cost over the datasets that we consider.

\medskip
\noindent\textbf{Experiment Setup:}
We consider groups of size $N=6$. In each experiment, we obtain the average performance of our algorithms by averaging over 50 {\em group trials}. 
For each group trial we pick a group of size $6$ according to a specific selection criterion, and we compute the performance of the different heuristics for this group. 
In particular, we evaluate the performance in two different {types of} network, namely:

\begin{enumerate}
 \item \textbf{Symmetric Configurations:} 
Users in the group have approximately the same expected number of users to meet among the group. Note that this is a relaxed requirement of symmetry with respect to the one used in Section \ref{sec::symmetric} where all the users were assumed to meet with the exact same probability.
 \item \textbf{Asymmetric Configurations:} 
Users in the group have different expected number of users to meet.
\end{enumerate}

\begin{figure*}
\centering
\begin{minipage}{1\textwidth}
 \subfigure[Direct Sharing - Asymm. Groups.]{
  \includegraphics[width=0.23\textwidth]{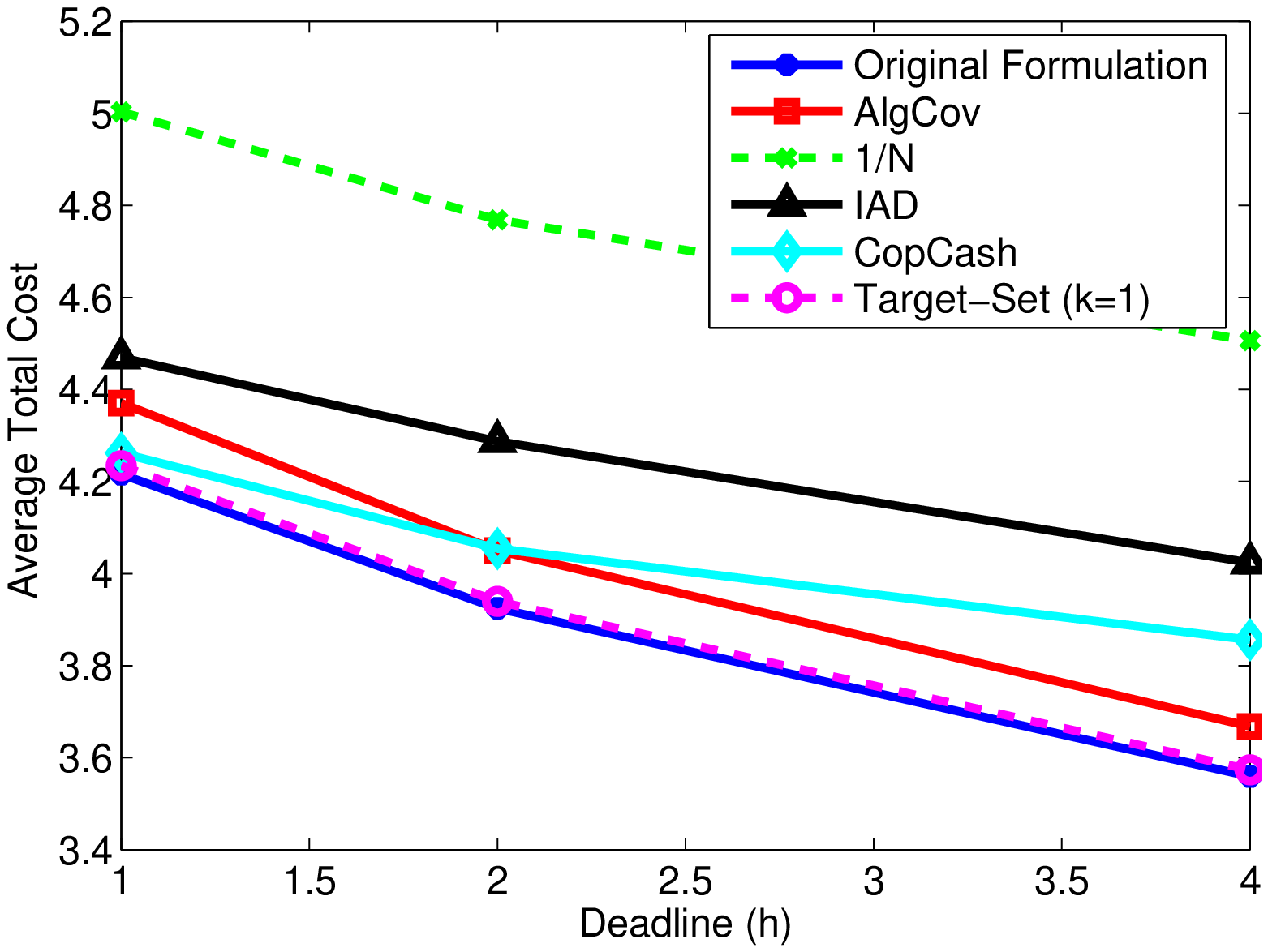}
  \label{fig::NoRouting_Asym}
  }
  \subfigure[Direct Sharing - Symm. Groups.]{
  \includegraphics[width=0.23\textwidth]{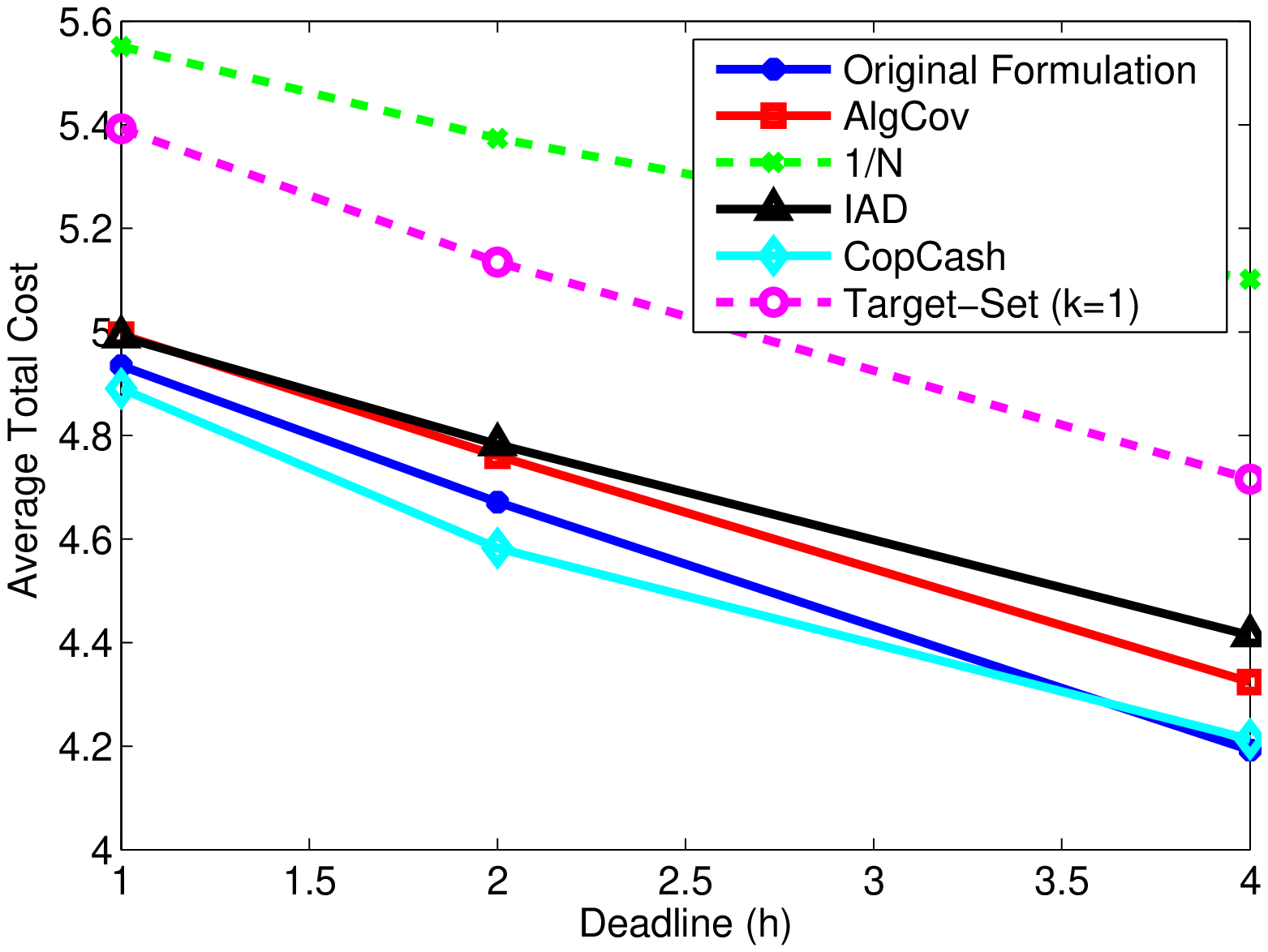}
  \label{fig::NoRouting_Sym}
  }
  \subfigure[Indirect Sharing - Asymm. Groups.]{
  \includegraphics[width=0.23\textwidth]{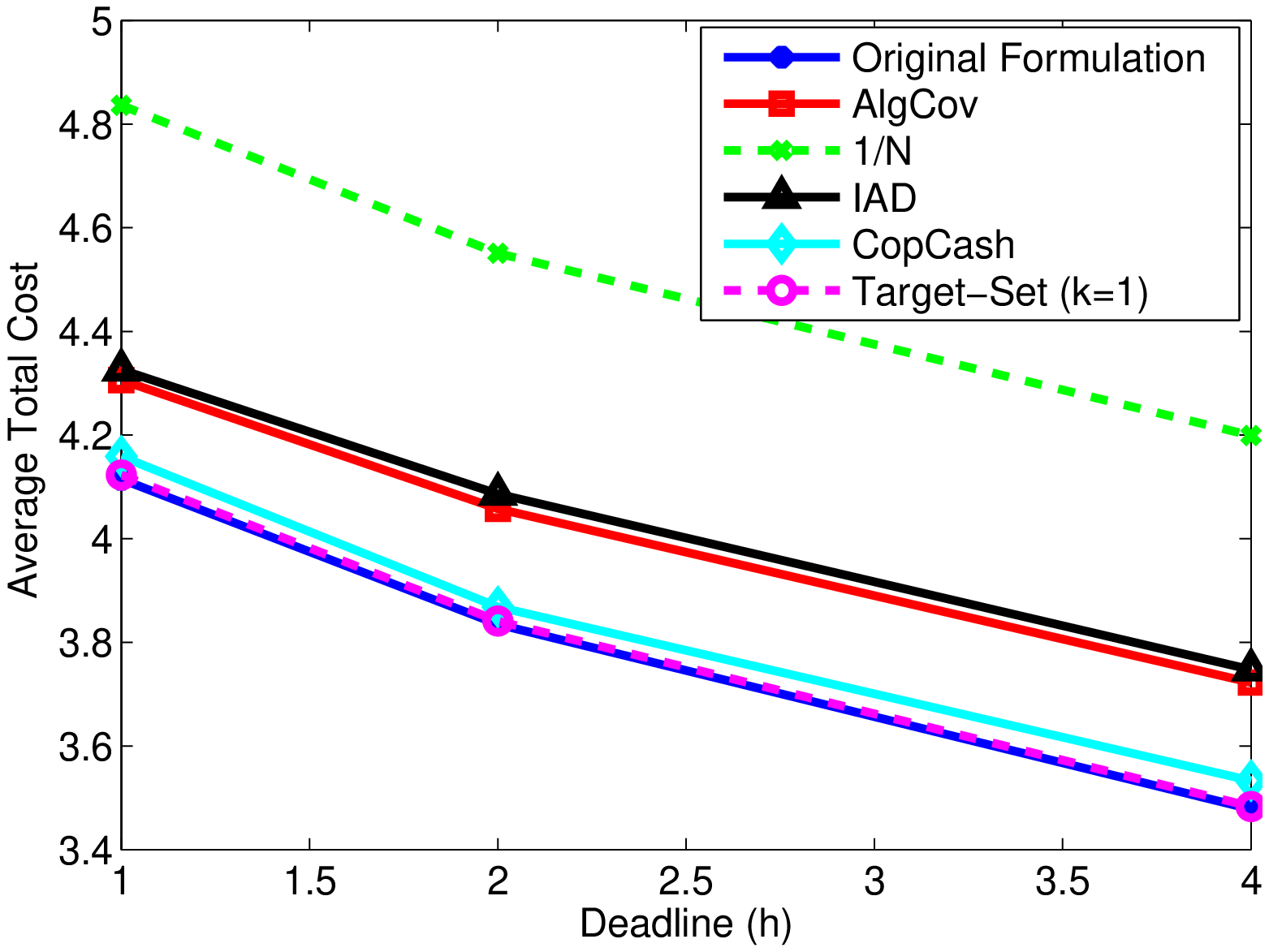}
  \label{fig::Routing_Asym}
  }
  \subfigure[Indirect Sharing - Symm. Groups.]{
  \includegraphics[width=0.23\textwidth]{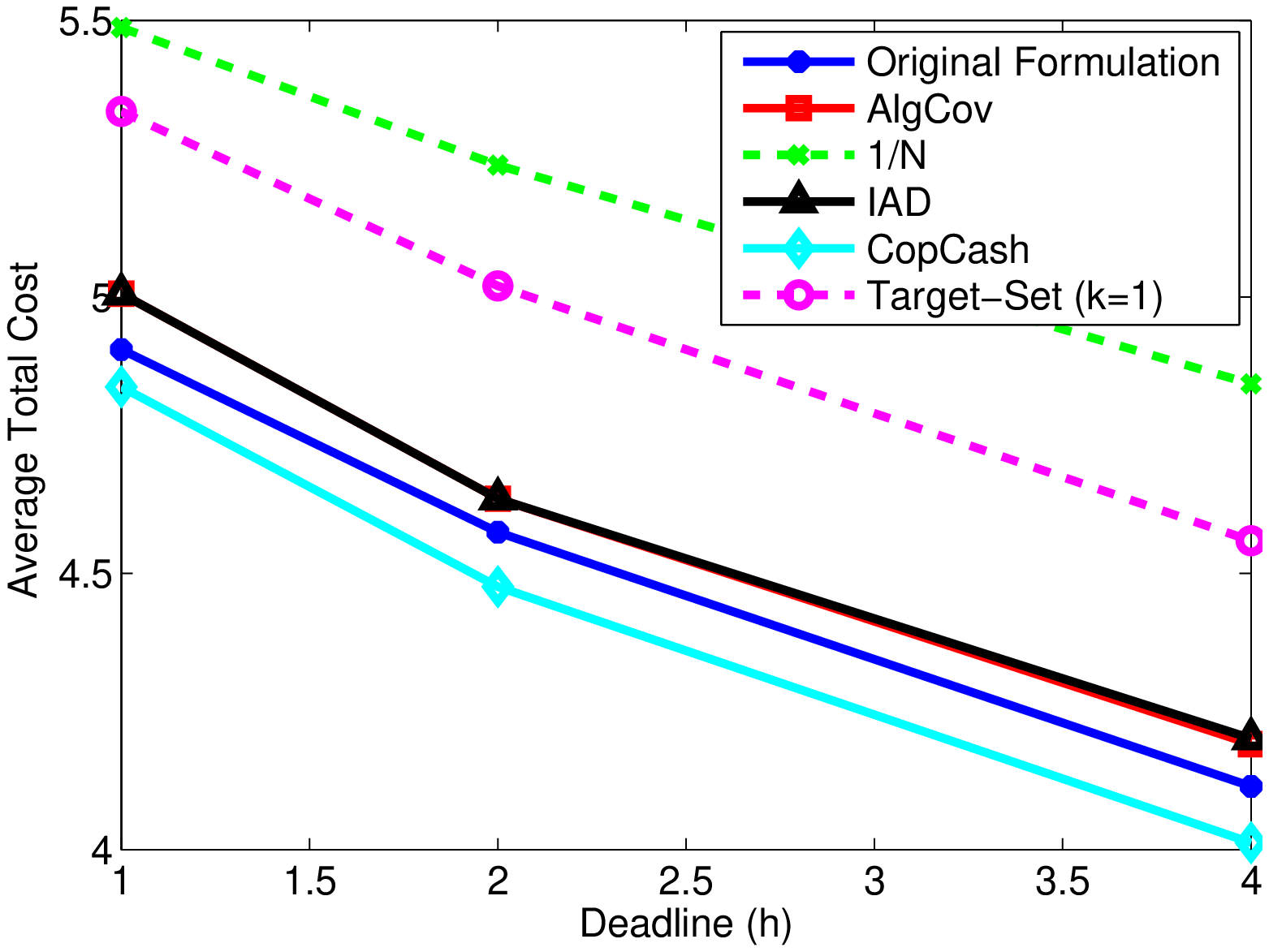}
  \label{fig::Routing_Sym}
 }
\vspace{-1mm}
\end{minipage}
\caption{Experimental result obtained from the MIT Reality Mining dataset.}
\label{fig::MIT}
\vspace{-3mm}
\end{figure*}

For each group, we define the {\em Expectation Deviation {($\text{ED}$):}} the difference between the maximum and the minimum expected number of encountered users, among all users, i.e., let $\mathcal{S}_j$ be the set of $N$ users belonging to group $j$, then
$\text{ED} = \max\limits_{i \in \mathcal{S}_j} \mathbb{E}(C_i) - \min\limits_{i \in \mathcal{S}_j} \mathbb{E}(C_i)$.
A group with high ED is more likely to have an asymmetric structure, while a group with small ED would have a symmetric structure. Our selection criterion is {therefore the following:}
(a) for asymmetric configurations, we choose groups that have $\text{ED} \geq th_{\text{asym}}$, and 
(b) for symmetric configurations, we select groups that have $\text{ED} \leq th_{\text{sym}}$, while having $\max\limits_{i \in \mathcal{S}_j} \mathbb{E}(C_i) \geq th_{\text{max}}$; 
$th_{\text{asym}}$, $th_{\text{sym}}$ and $th_{\text{max}}$ are decision parameters. 
For each experiment, these thresholds are set to values, which ensure the existence of the required number of groups.

We consider different \textit{deadlines} $t$: the time period after which all users must individually have the whole set of files at their disposal. 
Intuitively, we expect that the longer the deadline is, the higher the number of sharing opportunities can be among the users within the same group and thus the smaller the average cost  becomes.
For each deadline, the duration of the whole experiment is divided into a number of \textit{deadline trials}: for example, if the experiment is performed for a duration of $100$ days, and we consider a duration of $4$ hours in each day, then for a deadline of $2$ hours, we have $\frac{100 \cdot 4}{2} = 200$ deadline trials.


\subsection{MIT Reality Mining Dataset}
\label{sec::MIT}

\begin{figure}[t]
\begin{minipage}{0.5\textwidth}
\centering
\begin{tabular}{|l|c|}
\hline
\textbf{Description:} & \textbf{Value:} \\
\hline
Number of Users & $75$ \\
Dates of Interest & Oct., Nov., Dec. - 2004 \\
Hours of Interest & 2 pm - 6 pm \\
Sharing Period & $15$ mins \\
Deadlines $t$ & $1$ h, $2$ h, $4$ h \\
Number of Group Trials & $50$\\
$N$ & 6 \\
$th_{\text{asym}}$, $th_{\text{sym}}$, $th_{\text{max}}$ & $1.3$, $0.2$, $1.2$ \\
\hline
\end{tabular}
\vspace{0.1in}
\captionof{table}{Experiment Parameters - MIT Reality Mining Dataset.}
\label{exp1_details}
\vspace{-6mm}
\end{minipage}
\vspace{-2mm}
\end{figure}

We evaluate the performance of our proposed solutions and algorithms using the dataset from the MIT Reality Mining project \cite{pentland2009inferring}. Table \ref{exp1_details} lists the values of all the parameters that we use in our experiment, {described} in the following.

This dataset includes the traces from 104 subjects affiliated to MIT - 75 of whom were in the Media Laboratory - in the period from September 2004 to June 2005.
All subjects were provided with Nokia 6300 smart phones used to collect information such as the Bluetooth devices in the proximity logs.
In our experiment, we utilize this information to capture the sharing opportunities among users. 
Each device was programmed to perform a Bluetooth device discovery approximately every $5$ minutes, and store both the time instant at which the scan was performed, as well as the list of the MAC addresses of the devices found during the scan.

\medskip
\noindent\textbf{Assumptions:} We say that two users are connected at a time instant, if there exists a scan (at that time instant) that was performed by any of the two users, in which the other user was found. 
We assume {\it instantaneous} sharing, i.e., if two users are connected at a time instant, then they can share their full cache contents. We justify this assumption in the following discussion.
As specified in \cite{pentland2009inferring}, Bluetooth discovery scans were performed and logged approximately every $5$ minutes.
However, this granularity in time was not always attainable since (i) the devices were highly asynchronous, and 
(ii) some devices were powered off for a considerable amount of time.
Because a non-negligible fraction of users experienced these irregularities, discarding their traces is not a suitable solution.
Other solutions in the literature (for example, see \cite{chaintreau2007impact}) utilize the IDs of the cell towers to which mobile devices are connected to infer proximity information. 
However, such approaches are too optimistic in assuming sharing opportunities, and hence are not suitable for our application.
Our approach to deal with this highly irregular data was to consider the minimum sharing interval to be $15$ minutes, i.e.,
two users are connected for an entire sharing interval if they are so at any time instant in that specific interval.
Using the standard Bluetooth wireless transmission speed, this time period is sufficient to share approximately $2$ GBs of data. 
Hence, for all practical purposes, it is reasonable to assume that any two connected users can share their full cache contents during that sharing interval.

{For indirect sharing,} we do not allow {\em intra-interval relaying}: users cannot indirectly share with other users within the same interval. 
We do, however, allow {\em inter-interval relaying}: indirect sharing can be performed across successive intervals. 
Our premise is that, while a $15$-minute sharing interval is sufficient for one full cache content sharing, it {might} not be long enough to ensure more than one successful data exchange.
{This approach might severely limit the performance, i.e., a lower cost could be achieved by allowing intra-interval relaying.}

\begin{figure*}[t]
\centering
\begin{minipage}{1\textwidth}
\subfigure[Direct Sharing - Asymm. Groups.]{
 \includegraphics[width=0.23\textwidth]{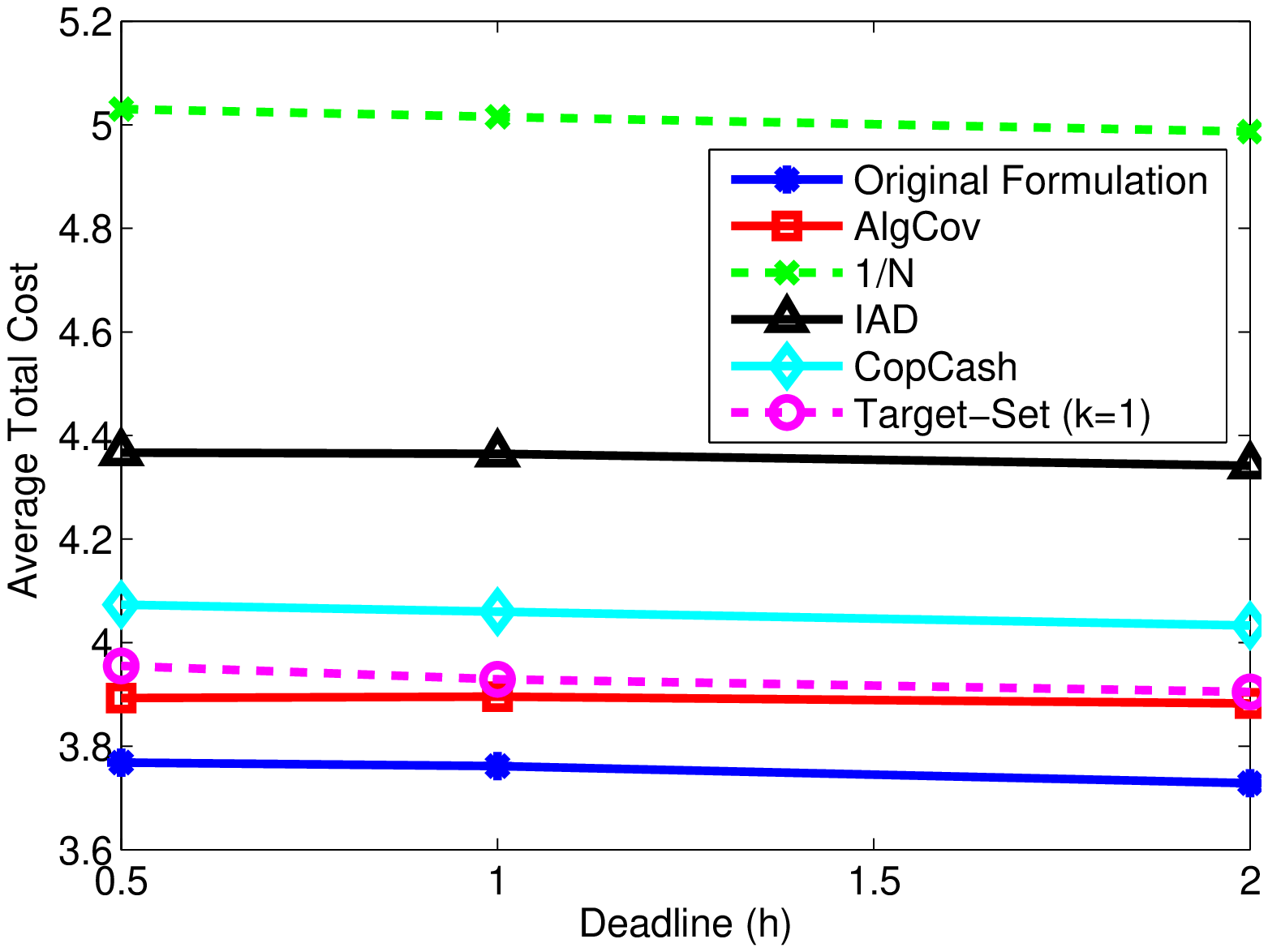}
   \label{fig::NoRouting_Asym_Infocom}
 }
 \subfigure[Direct Sharing - Symm. Groups.]{
 \includegraphics[width=0.23\textwidth]{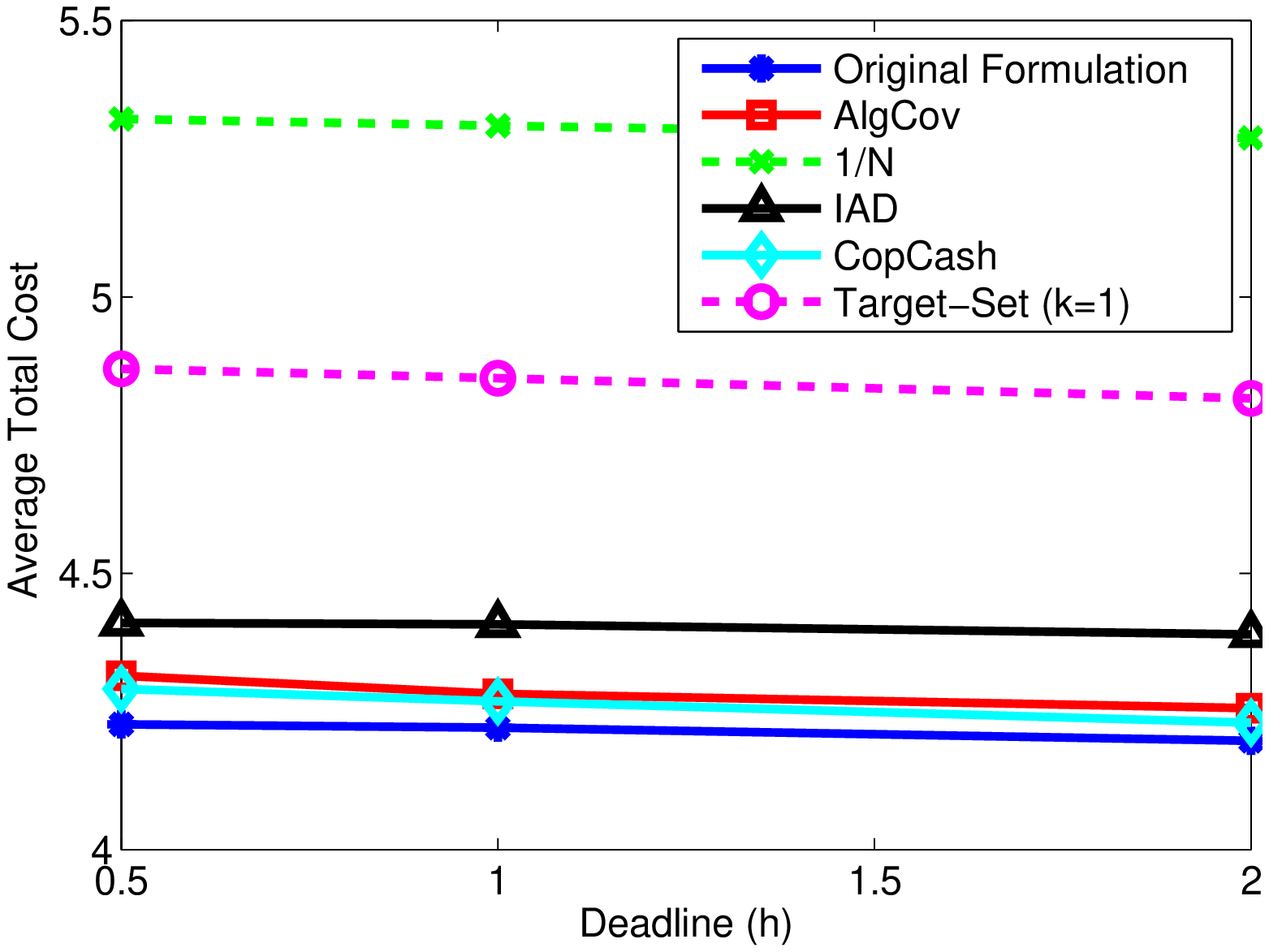}
   \label{fig::NoRouting_Asym_Infocom}
  }
  \subfigure[Indirect Sharing - Asymm. Groups.]{
  \includegraphics[width=0.23\textwidth]{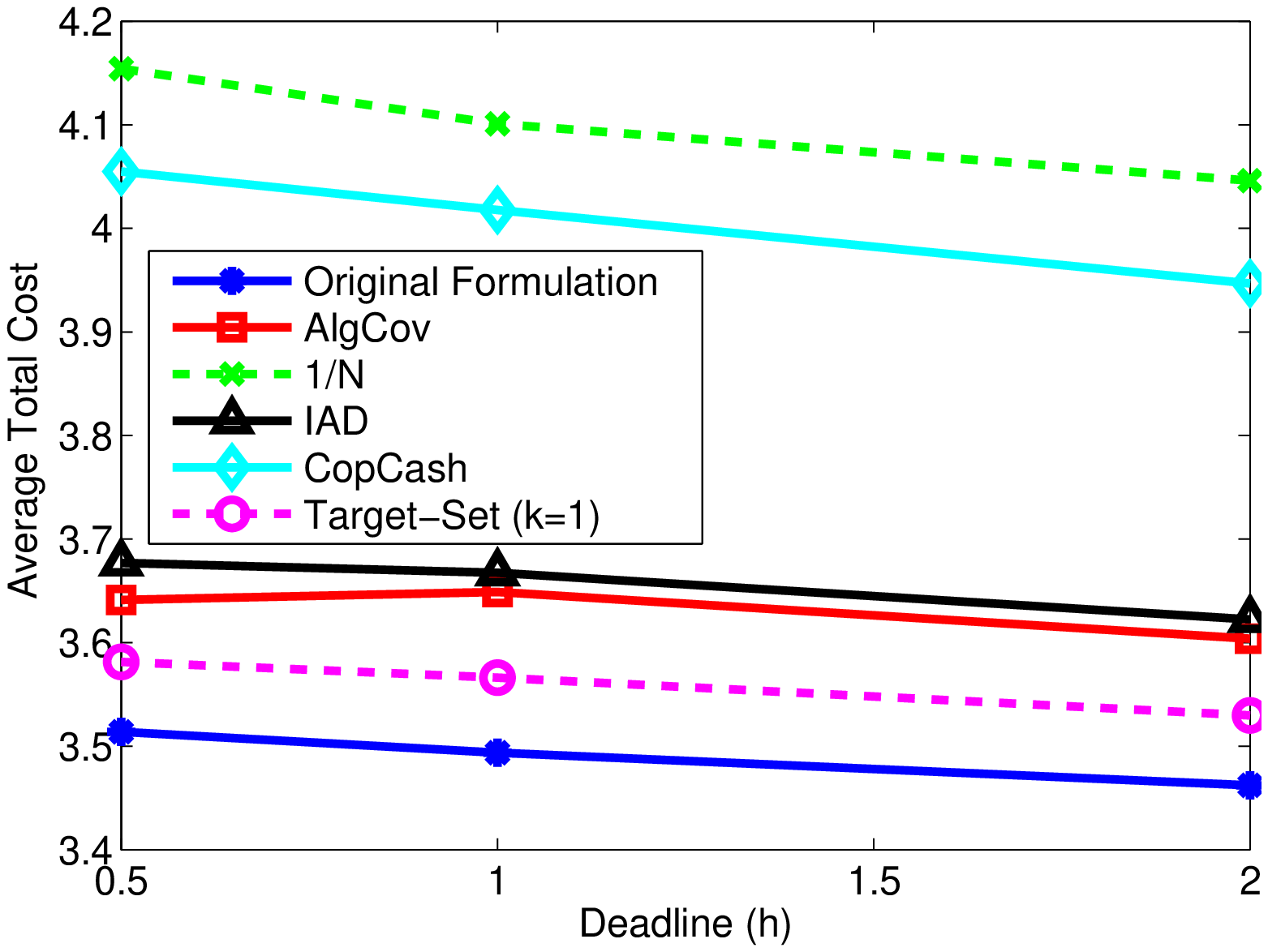}
  \label{fig::Routing_Asym_Infocom}
}
  \subfigure[Indirect Sharing - Symm. Groups.]{
  \includegraphics[width=0.23\textwidth]{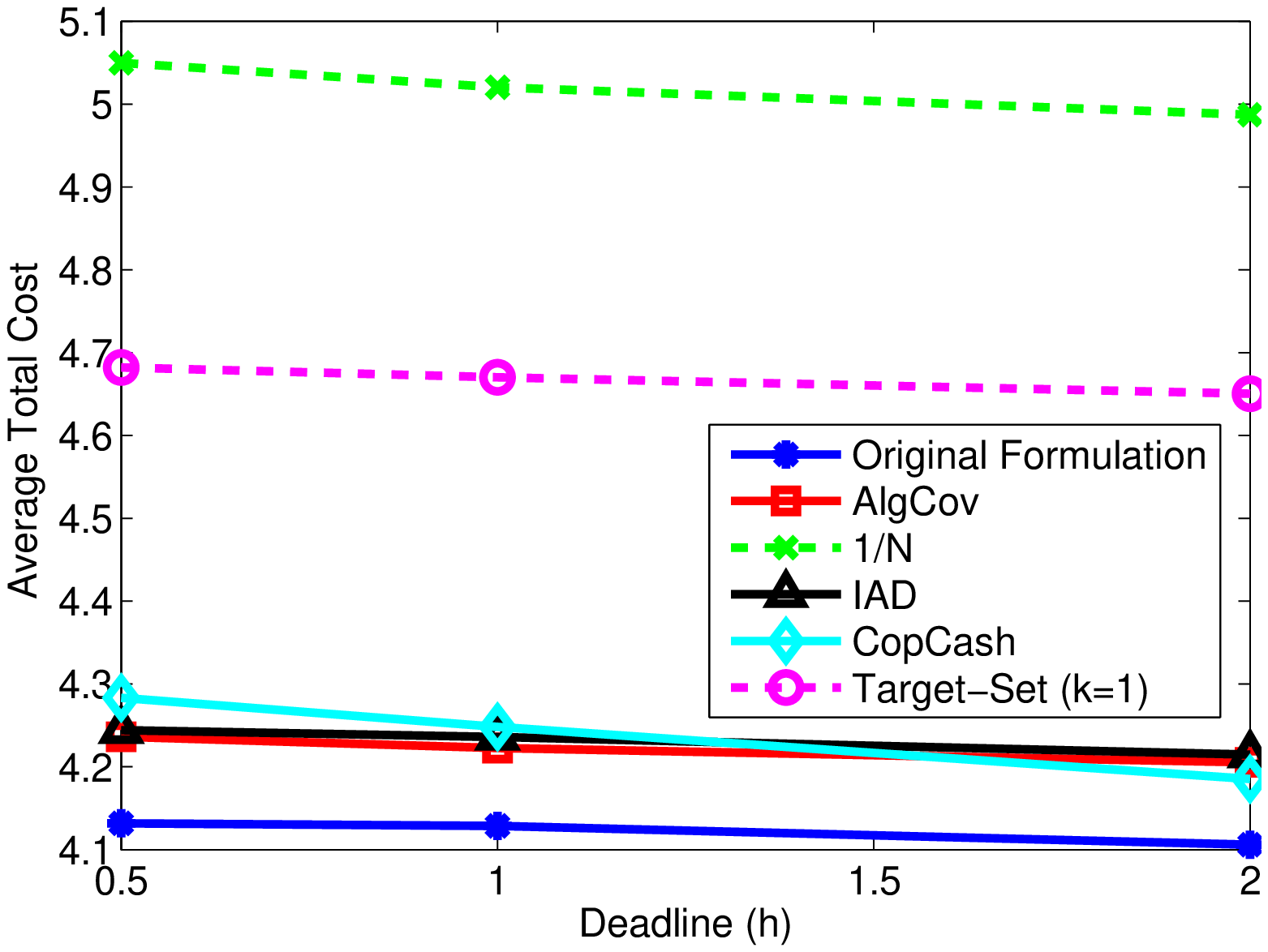}
  \label{fig::Routing_Sym_Infocom}
}
\vspace{-1mm}
\end{minipage}
\caption{Experimental results obtained from the synthesized Infocom-2005 trace.}
\label{fig::Infocom}
\vspace{-3mm}
\end{figure*}

\medskip
\noindent{\bf Setup:}
We consider a period of three months from the academic year 2004/2005 in MIT, namely from October to December. 
We consider {traces} of only 75 users - 
{labeled as}
affiliated to the Media Laboratory - during Monday, Tuesday and Wednesday.
 The reason {for} choosing these particular days is that we observed that, across the time period of interest, meetings occur most frequently in these days; thus, this represents a suitable period to assess the performance of all the solutions under consideration.
We perform each experiment from 2 pm to 6 pm, and we consider deadlines of $t \in \{1,2,4\}$ hours.
The thresholds for choosing groups are $th_{\text{asym}}=1.3$, $th_{\text{sym}} = 0.2$ and $th_{\text{max}} = 1.2$.
{The reason behind this particular choice was to ensure the existence of $50$ groups of $6$ users in the duration of the experiment.}

\medskip
\noindent{\bf Experimental Results:}
Figure \ref{fig::MIT} shows the performance of {different} network structures (i.e., asymmetric and symmetric) for the direct and indirect sharing cases, respectively.
From Figure \ref{fig::MIT}, as expected, we observe that: (i) the average total cost decreases as the deadline increases; (ii)
the average total cost in the indirect sharing case is less than the one in the direct case, thanks to a higher number of sharing opportunities;
(iii) using $1/N$ as a caching strategy performs the worst among all other schemes. 
This is because the $1/N$ scheme, differently from the other strategies, is not based on the meeting probabilities of the users.

\noindent
\underline{Asymmetric Networks:} Figure \ref{fig::NoRouting_Asym} and Figure \ref{fig::Routing_Asym} show the performance over asymmetric networks for the direct and {indirect} sharing cases, respectively. We note the following: \\
\noindent $\bullet$ Target-Set performs very close to the optimal scheme in both the direct and the indirect sharing cases. 
This is due to the asymmetric structure of the selected groups: one node is more likely to be connected to the other members of the group, and therefore the optimal solution would rely on that node to deliver the data to the whole group.\\
\noindent $\bullet$ AlgCov outperforms {IAD} in Figure \ref{fig::NoRouting_Asym}, which indicates that AlgCov utilizes the solution that is generated from {PSC}. In contrast, AlgCov and the IAD strategy perform almost the same in Figure \ref{fig::Routing_Asym} which indicates that IAD outperforms {PSC} in this case. 
This justifies the merge between these two heuristics {in the design of AlgCov}.

\noindent\underline{Symmetric Networks:}
Figure \ref{fig::NoRouting_Sym} and Figure \ref{fig::Routing_Sym} show the performance of the different schemes over symmetric networks for the direct and the indirect sharing cases, respectively. 
Observations are similar to those drawn for the asymmetric case. However, one major observation is that {Target-Set}, differently from asymmetric groups, poorly performs.
This is a direct consequence of the symmetric structure of the selected group: in a symmetric group, an optimal sharing strategy would equally distribute the caching and sharing efforts among
all members within the {group; in contrast, {Target-Set} selects only one member who has the task of caching and sharing the data for the group.}

\begin{remark}
One might argue that CopCash has an inherent advantage over the other caching strategies since it does not need the genie-aided information of the pairwise meeting probabilities. 
{However, this information is not hard to obtain in a realistic scenario. For example, although being out of the scope of this work, one can think of modifying AlgCov, by including a learning module. With this and by exploiting the regular mobility behavior of the users, the probabilities can be estimated as reportedly done in the literature (see \cite{barbera2014data,chaintreau2007impact}).}
\end{remark}
\begin{remark}
CopCash performs closely to our proposed solution. 
One can thus draw a premature conclusion that pre-caching does not bring significant benefits with respect to opportunistically exploiting sharing opportunities, as CopCash does. 
This is true when the meeting probabilities are small, as in the MIT Reality Mining dataset.
However, as shown next, pre-caching solutions outperform opportunistic sharing approaches when the users are moderately/highly connected. 
\end{remark}
\begin{figure}
\begin{minipage}{0.5\textwidth}
\centering
\begin{tabular}{|l|c|}
\hline
\textbf{Description:} & \textbf{Value:} \\
\hline
Number of Users & $300$ \\
Duration of Experiment & {$3$ (Infocom-2005) and} \\
 & { $11$ (Cambridge-2006) days} \\
Sharing Period & $10$ mins \\
Deadlines $t$ & $0.5$ h, $1$ h, $2$ h \\
Number of Group Trials & $50$\\
$N$ & 6 \\
$th_{\text{asym}}$, $th_{\text{sym}}$, $th_{\text{max}}$ & $3.7$, $0.2$, {$1.2$} \\
\hline
\end{tabular}
\vspace{0.1in}
\captionof{table}{Experiment Parameters - Infocom-2005 and Cambridge-2006.}
\label{exp2_details}
\vspace{-6mm}
\end{minipage}
\vspace{-2mm}
\end{figure}

\begin{figure*}[t]
\centering
\begin{minipage}{1\textwidth}
\subfigure[Direct Sharing - Asymm. Groups.]{
 \includegraphics[width=0.23\textwidth]{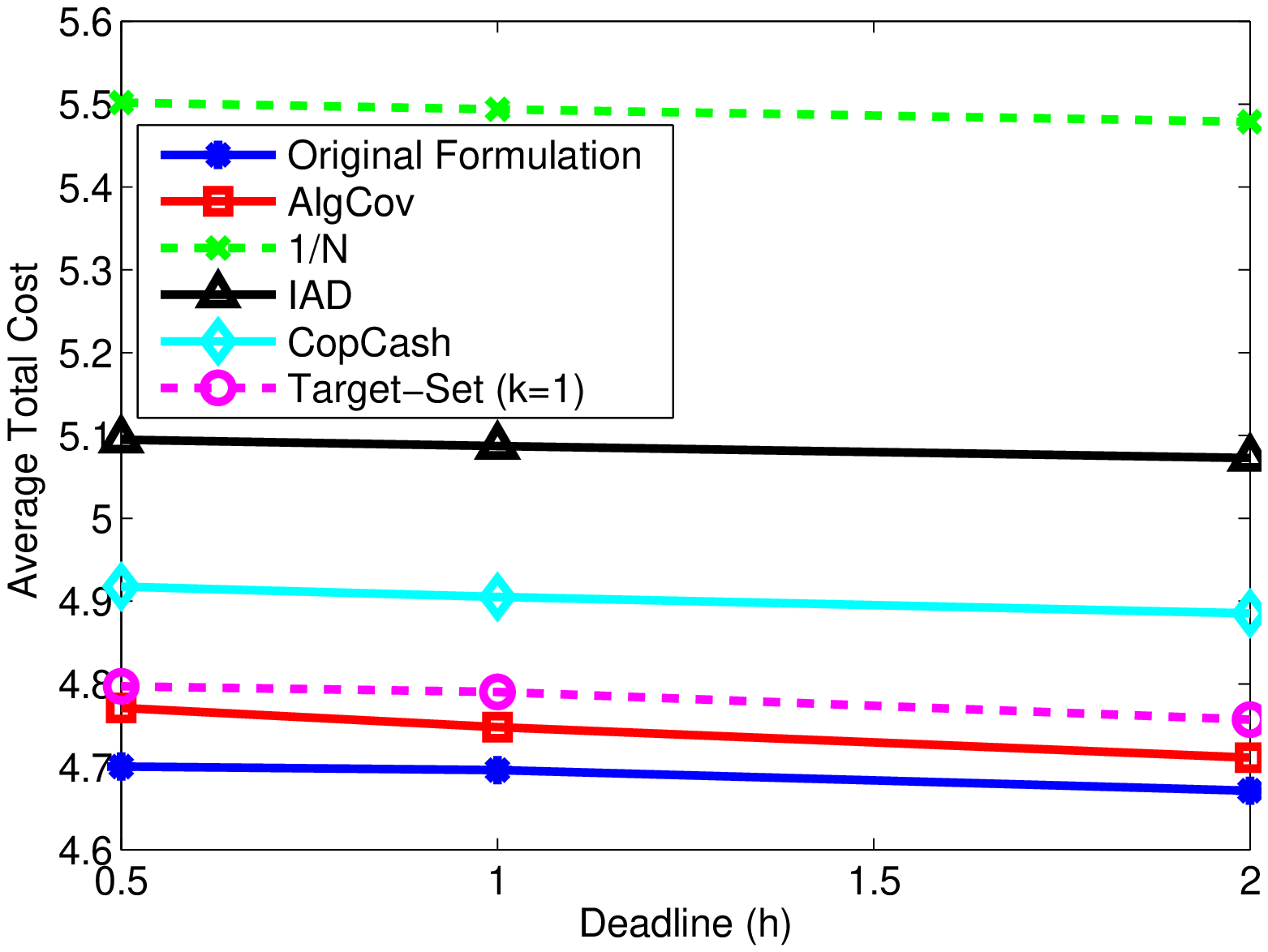}
   \label{fig::NoRouting_Asym_Cambridge}
 }
 \subfigure[Direct Sharing - Symm. Groups.]{
 \includegraphics[width=0.23\textwidth]{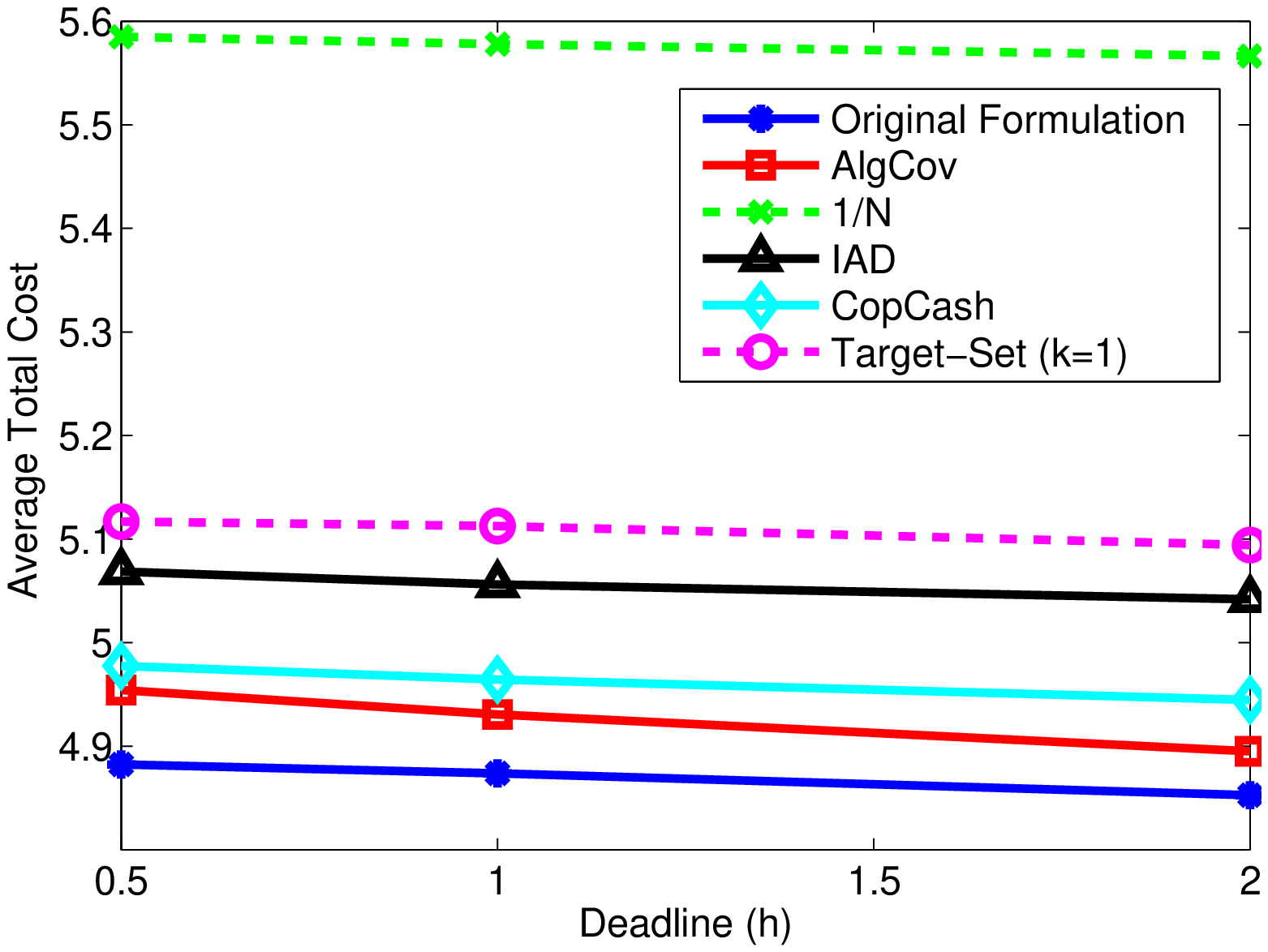}
   \label{fig::NoRouting_Asym_Cambridge}
 }
  \subfigure[Indirect Sharing - Asymm. Groups.]{
  \includegraphics[width=0.23\textwidth]{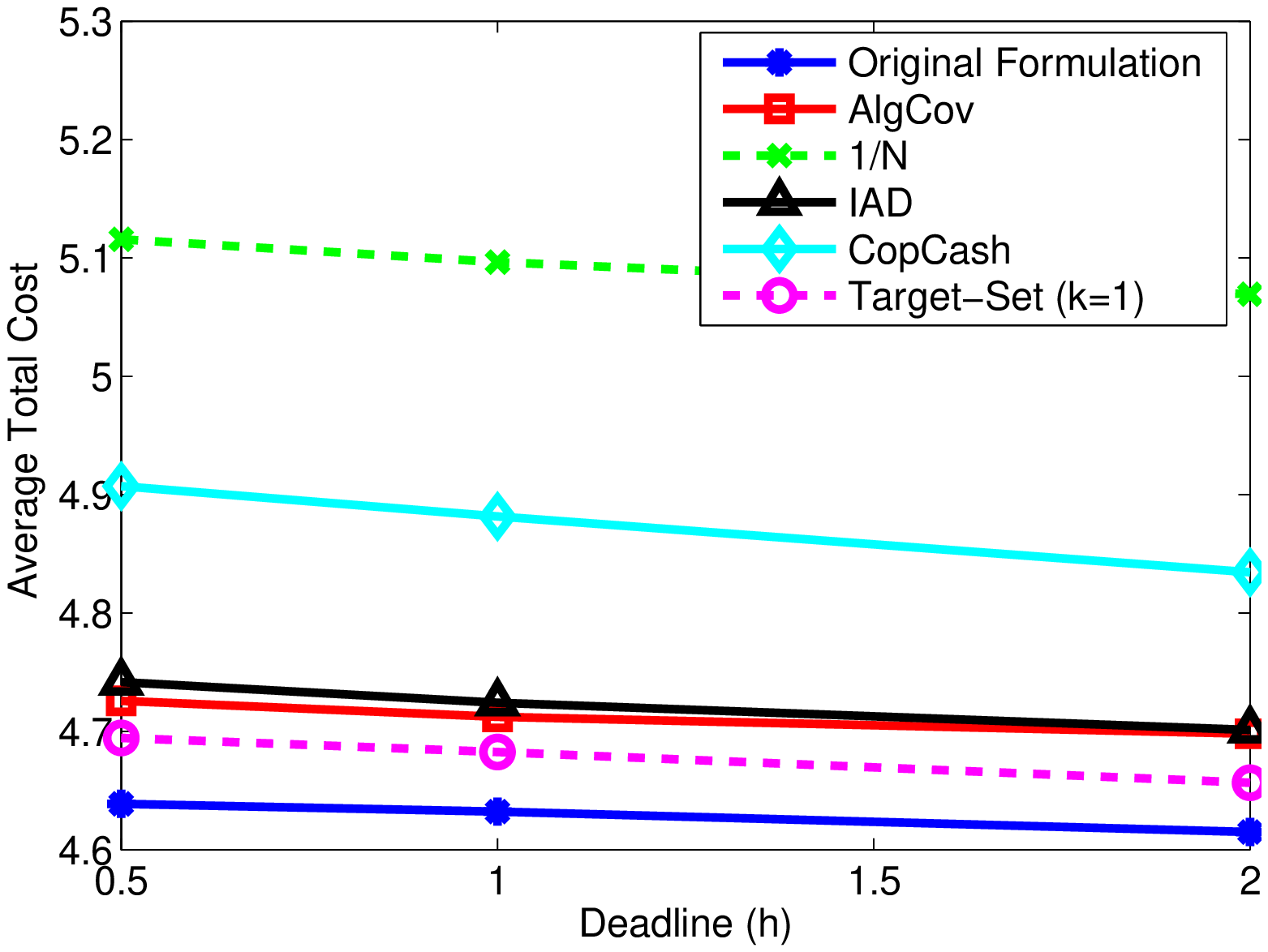}
  \label{fig::Routing_Asym_Cambridge}
}
  \subfigure[Indirect Sharing - Symm. Groups.]{
  \includegraphics[width=0.23\textwidth]{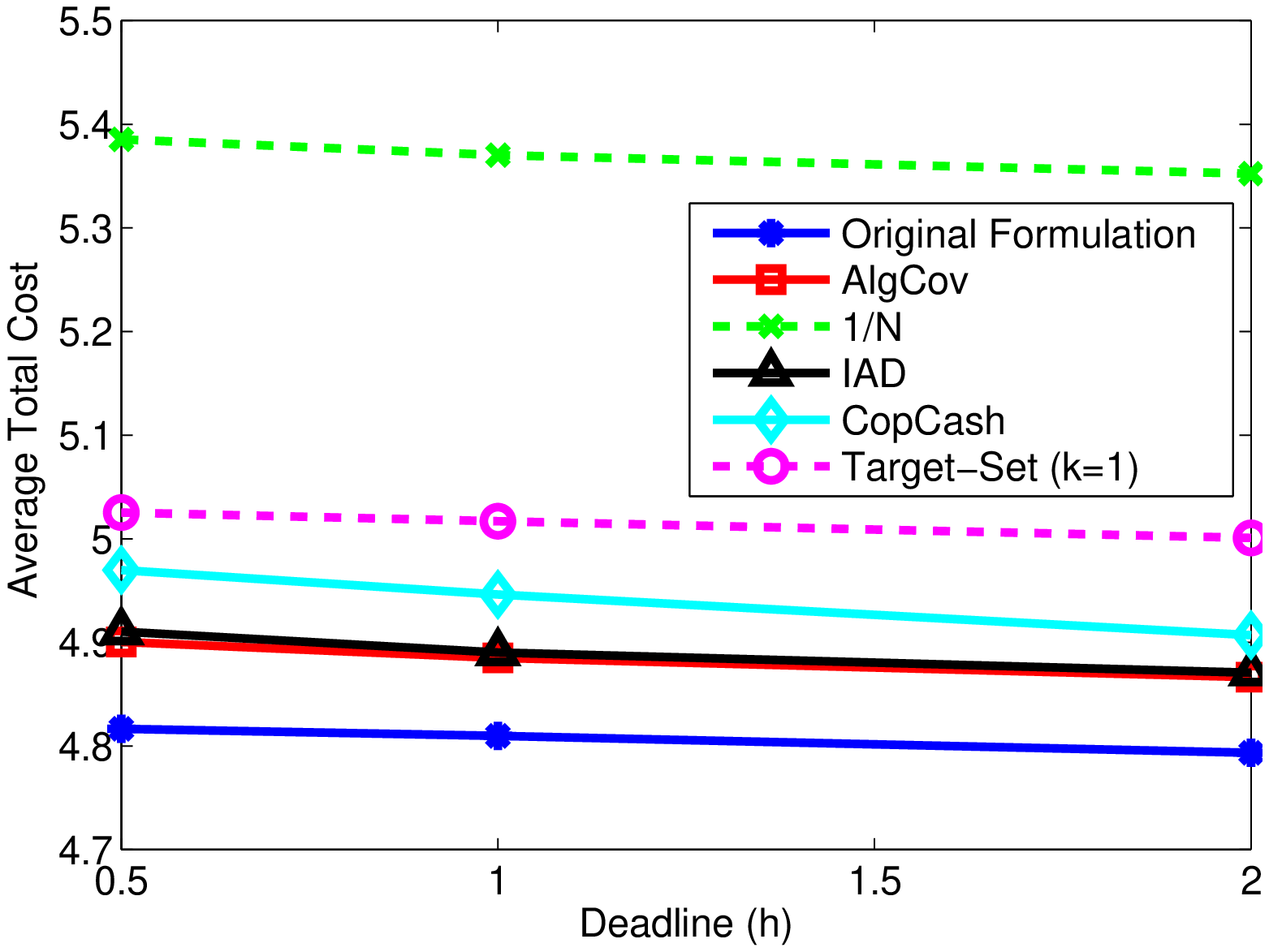}
  \label{fig::Routing_Sym_Cambridge}
}
\vspace{-1mm}
\end{minipage}
\caption{results obtained from the synthesized Cambridge-2006 trace.}
\label{fig::Cambridge}
\vspace{-3mm}
\end{figure*}

\subsection{SWIM-Based Results}

We here evaluate the performance of our algorithms over mobility traces synthesized using the SWIM model. SWIM \cite{kosta2010small} is a human mobility model that is used to synthesize mobility traces based on the users social behavior.
Traces are generated in the form of {\em events}: the exact time at which two users meet/leave. 
Thus, the trace files consist of a chronological series of meeting/leaving events among the users involved in the generation of the trace.
We use a synthesized version of two existing traces, namely Infocom-2005 and Cambridge-2006. 
These traces were obtained {through} experiments conducted in the IEEE INFOCOM 2005 conference and in Cambridge in 2006, respectively ({see \cite{hui2005pocket,chaintreau2005pocket} for more details)}. 
The synthesized versions of these traces include a greater number of nodes (with the same spatial density) than the original ones, {which is the main reason behind our choice of the synthesized traces.}

\medskip
\noindent\textbf{Assumptions:} We consider the sharing interval to be $10$ minutes. We say that 
two users successfully exchange their cache contents if they are in contact for at least $85\%$ of the interval.
Similarly to Section \ref{sec::MIT}, in the indirect sharing we 
only allow inter-interval relaying.

\medskip
\noindent{\bf Setup:}
We perform each experiment over the traces from $300$ virtual users during the entire duration of the trace ($3$ days for Infocom-2005 and $11$ days for Cambridge-2006). 
The deadlines that we consider are of $t=0.5$ hour, $t=1$ hour and $t=2$ hours.
The thresholds for choosing groups are $th_{\text{asym}}=3.7$, $th_{\text{sym}} = 0.2$ and $th_{\text{max}} = 1.2$. 
{The reason behind this particular choice was to ensure the existence of $50$ groups of $6$ users for all the days of the experiment.}
Table \ref{exp2_details} lists the values of all the parameters of the experiments.

\medskip
\noindent{\bf Experimental Results:}
We assess the performance of our algorithms on the Infocom-2005 (Figure \ref{fig::Infocom}) and Cambridge-2006 (Figure \ref{fig::Cambridge}) mobility traces. 
Similar conclusions to those in Section \ref{sec::MIT}
can be drawn. In particular: (i) the average total cost decreases as the deadline increases; (ii) the average total cost incurred in the indirect sharing case is less than the one in the direct counterpart; (iii) the caching strategy $1/N$ shows the worst performance among the different schemes; 
(iv) {Target-Set} performs close to the optimal in asymmetric configurations.
However, differently from Section \ref{sec::MIT},
in most of the cases CopCash poorly performs with respect to other solutions.
The reason is that the mobility traces of both Infocom-2005 and Cambridge-2006 show a relatively high frequency of meetings among users, which is a distinct feature with respect to the MIT Reality Mining dataset.

\section{Conclusions}  \label{sec:concl}
We here motivated, proposed, analysed, and experimentally evaluated AlgCov,
a simple low-complexity algorithm for social caching, that uses pre-caching in 
anticipation of encounter opportunities to minimize the required download bandwidth.
We derived formal LP formulations and presented a worst-case analytical performance gap.
We numerically evaluated the performance of the proposed solutions on (i) the mobility traces obtained from the MIT Reality Mining data set, and (ii) two mobility traces that were synthesized using the SWIM mobility model. 
AlgCov achieves a performance which is close to the optimal and, in some configurations, it outperforms existing solutions, such {as} the {Target-Set}.
AlgCov makes the case that, even in the presence of random encounters, {using} simple algorithms for pre-caching can significantly reduce bandwidth usage.

\appendices

\section{Proof of Theorem \ref{theorem_1}}
\label{app:proofOfLPreduction}

The key observation is to notice that the constraints in the LP in \eqref{opt_problem_original_no_relay} can be written in the form 
$1 - \pi_v \mathbf{x} \leq y_{i,k}, \ \forall y_{i,k} \in \mathcal{Y}^{(v)}$,
where $\mathcal{Y}^{(v)} = \{y_{i,k} | 1 - \pi_v \mathbf{x} \leq y_{i,k} \text{ is a constraint} \}$.
Since all the constraints of the type $\mathbf{1}_N - \mathbf{A}^{(k)} \mathbf{x} \leq \mathbf{y}_k$ in the LP in \eqref{opt_problem_original_no_relay} can be replaced with $1 - \pi_v \mathbf{x} \leq y_{i,k}$, the optimal solution would make all $y_{i,k} \in \mathcal{Y}^{(v)}$ equal, as proved in Lemma \ref{lemma_th1}.
\begin{lemma}
\label{lemma_th1}
Let ($\hat{\mathbf{x}}$,$\hat{\mathbf{y}}$) be an optimal solution for the LP in \eqref{opt_problem_original_no_relay}. Then $\hat{y}_{i,k} = y_{v},\ \forall \hat{y}_{i,k} \in \mathcal{Y}^{(v)}$. Thus, 
\begin{align*}
f^{\text{Opt}}(\hat{\mathbf{x}},\hat{\mathbf{y}}) = \sum\limits_{v = 1}^{2^N - 1} y_{v} \sum\limits_{i=1}^{N} \sum_{\substack{k=1, \\ y_{i,k} \in \mathcal{Y}^{(v)}}}^K p_k + \mathbf{1}^T_N \hat{\mathbf{x}}.
\end{align*}
\end{lemma}
We next prove the result in Lemma \ref{lemma_th1}.
Without loss of generality, assume that $\hat{y}_{i,k} = y_{v},\ \forall \hat{y}_{i,k} \in \mathcal{Y}^{(v)} \backslash \bar{y}$ and $\bar{y} = y_{v} + \Delta$ where $\Delta \geq 0$. 
Then, since this is a feasible point, $\Delta$ can be driven down to zero without violating the feasibility conditions, and consequently reducing the optimal value of the objective function; thus, we have a contradiction. {The same argument can be extended} to the case where more than one $\hat{y}_{i,k}$ is different from $y_{v}$.

{With $y_{i,k} \! = \! y_{v},\ \forall y_{i,k} \! \in \! \mathcal{Y}^{(v)}$ in $f^{\text{Opt}}(\mathbf{x},\mathbf{y})$ {in \eqref{opt_problem_original_no_relay}}, we get}
\begin{equation*}
\begin{split}
\sum\limits_{k=1}^{K} p_k \sum\limits_{i=1}^N y_{i,k} &= \sum\limits_{k=1}^{K} p_k \sum\limits_{i=1}^N \sum_{v=1}^{2^N-1} y_{i,k} \mathbb{1}_{\{y_{i,k} \in \mathcal{Y}^{(v)} \}}\\
&= \sum\limits_{v = 1}^{2^N - 1} y_{v} \sum\limits_{i=1}^{N} \sum_{k=1}^K p_k \mathbb{1}_{\{y_{i,k} \in \mathcal{Y}^{(v)} \} }.
\end{split}
\end{equation*}
This concludes the proof of Lemma \ref{lemma_th1}.

Notice that, by our definition in Theorem \ref{theorem_1}, we have $\sum\limits_{i=1}^{N} \sum_{k=1}^K p_k \mathbb{1}_{\{y_{i,k} \in \mathcal{Y}^{(v)} \} } = \sum_{u \in \mathcal{S}_v} \text{Pr}(u \rightarrow \mathcal{S}_v).
$.

We now use the result in Lemma \ref{lemma_th1} to prove Theorem \ref{theorem_1}, i.e., the equivalence of the LPs in \eqref{opt_problem_original_no_relay} and in \eqref{opt_problem_original_no_relay_simplified}.
%
\\ {\bf Part 1.} Let $ \left(\mathbf{x}^1,\mathbf{y}^1 \right) $ be an optimal solution for the LP in \eqref{opt_problem_original_no_relay}, which follows the structure described in Lemma \ref{lemma_th1}.
For $v \in [1:2^N-1]$, let $y_{v}^1$ be the value where, for each $y^{1}_{i,k} \in \mathcal{Y}^{(v)}$, $y^{1}_{i,k} = y_{v}^1$.
Then, one can construct a feasible solution $ \left(\mathbf{x}^2,\mathbf{y}^2 \right)$ for the LP in \eqref{opt_problem_original_no_relay_simplified} as follows: (i) set ${\mathbf{x}^2 = \mathbf{x}^1}$; (ii) let {$y_{v}^2$} be an element of $\mathbf{y}^2$ that corresponds to a constraint of the form $1 - \pi_v \mathbf{x}^2 \leq {y_{v}^2}$ in the LP in \eqref{opt_problem_original_no_relay_simplified}, then set $y_{v}^2 = y_{v}^1$.
By doing so, the constraints of the LP in \eqref{opt_problem_original_no_relay_simplified} are satisfied. Moreover, {with} Lemma \ref{lemma_th1}, the objective functions of both problems are equal, when evaluated at the described points.
\\ {\bf Part 2.}
Let $ \left(\mathbf{x}^2,\mathbf{y}^2 \right)$ be an optimal solution for the LP in \eqref{opt_problem_original_no_relay_simplified}. Then one can construct a feasible solution $\left(\mathbf{x}^1,\mathbf{y}^1 \right)$ for the LP in \eqref{opt_problem_original_no_relay} as follows: (i) set $\mathbf{x}^1 = \mathbf{x}^2$; (ii) {$\forall y_{i,k}^1 \in \mathcal{Y}^{(v)}$, set $y_{i,k}^1 = y_v^2.$ }.
By doing so, the constraints of the LP in \eqref{opt_problem_original_no_relay} are guaranteed to be satisfied. Moreover, {with} Lemma \ref{lemma_th1}, the objective functions of both problems will be equal, when evaluated at the described points. This concludes the proof.

\section{Proof of Theorem \ref{theorem_2}}
\label{app:SimplificationSymmetric}
We prove the equivalence of the LP in \eqref{opt_problem_original_no_relay} and the LP in \eqref{opt_problem_general_N_no_relay_simplified} by means of the following lemma.

\begin{lemma}
\label{lemma:optSol}
Let $\left(\hat{\mathbf{x}},\hat{\mathbf{y}} \right )$ be the optimal solution for the LP in \eqref{opt_problem_original_no_relay}. Then, by assuming a symmetric model
\begin{enumerate}
\item $\hat{\mathbf{x}} = \hat{x}\mathbf{1}_N$, with $\hat{x} \in [0,1]$; 
\item { For $m \in [1:N]$, let $\Pi^{(m)} = \{ (i,k) | \left [1 - m \hat{x}\right]^+ \leq \hat{y}_{i,k} \leq \left [1-(m-1)\hat{x}\right ]^+\} $ with $i \in [1:N]$, $k\in [1:K]$, then $\hat{y}_{i,k} = \hat{y}_{m}, \forall (i,k) \in  \Pi^{(m)}$, with $\hat{y}_m = \left [1-mx \right ]^+$.}
\end{enumerate}
Moreover, with reference to $f^{\text{Opt}}(\mathbf{x},\mathbf{y})$ in \eqref{opt_problem_original_no_relay}, we get
\begin{align*}
\sum\limits_{k=1}^{K} p_k \sum\limits_{i=1}^{N} \hat{y}_{i,k} = 
\sum\limits_{i=1}^{N} \hat{y}_i {N-1 \choose i-1} N p^{i-1} (1-p)^{N-i}.
\end{align*}
\end{lemma}

We now prove Lemma \ref{lemma:optSol} in three steps. \\
{\bf{Step 1}}.
We prove that $\mathbf{x} = x \mathbf{1}_N$ and {$y_{i,k} = y_m  \; \forall \; (i,k) \in \Pi^{(m)}, \; m \in [1:N]$} is a feasible solution for the LP in \eqref{opt_problem_original_no_relay}. 
Assume a feasible solution consists of $\mathbf{x} = x \mathbf{1}_N$. Then, with {$y_{i,k} \!=\! y_{m} \; \forall \; (i,k) \!\in\! \Pi^{(m)}, \; m \!\in\! [1:N] $ so that $y_{m} \!\geq\!  \left [ 1 \!-\! m x \right ]^+$}, we get a feasible solution of the required form.\\
{\bf{Step 2}}. Assume that an optimal solution has $\hat{\mathbf{x}} = \hat{x} \mathbf{1}_N$. 
We prove, by contradiction, that this implies {$\hat{y}_{i,k} = \hat{y}_{m}\; \forall \; (i,k) \in \Pi^{(m)}$. We use similar steps as in the proof of Lemma \ref{lemma_th1}.} 
Without loss of generality, assume that {$\hat{y}_{i,k} = \hat{y}_{m} \; \forall \; (i,k) \in \Pi^{(m)} \backslash (\bar{i},\bar{k})$ and $\hat{y}_{\bar{i},\bar{k}} =\hat{y}_m + \Delta$, where $\Delta \geq 0$.
Since this point is feasible, then $\Delta$ can be driven down to zero without having violated the feasibility conditions; this operation (i.e., setting $\Delta=0$) also implies a reduction in the optimal value of the objective function; thus we have a contradiction.
}
\\
{\bf{Step 3}}. We prove that, {for the symmetric model,} an optimal solution of the form 
$\hat{\mathbf{x}} = \hat{x} \mathbf{1}_N$ and {$\hat{y}_{i,k} = \hat{y}_{m} \; \forall \; (i,k) \in \Pi^{(m)} \; \forall m \in [1:N] $ exists.}
Without loss of generality, assume that $\left(\tilde{\mathbf{x}}, \tilde{\mathbf{y}} \right)$ is an optimal solution of the form {$\tilde{\mathbf{x}} = [x, \ \cdots, \ x, \ x+\Delta]^T$,} 
where $0 \leq \Delta < x$\footnote{This assumption is not necessary and is made only to simplify the analysis.}. 
We show that
$\hat{\mathbf{x}} = x + \frac{\Delta}{N} \mathbf{1}_N$
gives a smaller value for the objective function of the LP in \eqref{opt_problem_original_no_relay}, i.e., ${f}^{\text{Opt}}(\tilde{\mathbf{x}},\tilde{\mathbf{y}}) - {f}^{\text{Opt}}(\hat{\mathbf{x}},\hat{\mathbf{y}}) = \sum\limits_{k=1}^{K} p_k \sum\limits_{i=1}^N \tilde{y}_{i,k} - \sum\limits_{k=1}^{K} p_k \sum\limits_{i=1}^N \hat{y}_{i,k} \geq 0$.
We start by noticing that, by using the symmetric model in $f^{\text{Opt}}(\mathbf{x},\mathbf{y})$ in \eqref{opt_problem_original_no_relay}, we get
{{
\begin{align}
\label{eq11}
&\sum\limits_{k=1}^{K} p_k \sum\limits_{i=1}^N {y}_{i,k} = p_1 \left (\sum\limits_{i=1}^N {y}_{i,1} \right) + \hdots + p_K \left(\sum\limits_{i=1}^N {y}_{i,K} \right) \nonumber \\
& {=}  \sum\limits_{j=0}^{\frac{N(N-1)}{2}} \!\!B \left( p,j,N\right)  \sum\limits_{k=1}^K \sum\limits_{i=1}^N{y}_{i,k}
\mathbb{1}_{\left \{p_k = B \left( p,j,N\right)  \right \} }, 
\end{align}}
where the last equality follows by noticing that each of the $p_k, k\in[1:K]$ is equal to a term of the type $B \left( p,j,N\right) =p^j \left( 1-p \right)^{\frac{N(N-1)}{2}-j}$ for some $j\in \left [0:\frac{N(N-1)}{2} \right ]$ and by swapping the order of the summations.

We next evaluate ${f}^{\text{Opt}}(\tilde{\mathbf{x}},\tilde{\mathbf{y}})$ and ${f}^{\text{Opt}}(\hat{\mathbf{x}},\hat{\mathbf{y}})$ separately.
\\
\textbf{Evaluation of ${f}^{\text{Opt}}(\tilde{\mathbf{x}}, \tilde{\mathbf{y}})$:}
We define 
\begin{align*}
 \tilde{\Pi}^{(m,j)}_{\Delta} = & 
\left \{ (i,k) | p_k = B \left( p,j,N\right),  \right. 
\\
   &
\left.  \left [1-mx-\Delta \right ]^+ \leq \tilde{y}_{i,k} < \left [1-m x\right ]^+ \right \}, \\
 \tilde{\Pi}^{(m,j)}_{\Delta,\dagger} = & \left \{(i,k) | p_k = B \left( p,j,N\right), \right. \\
  & \left. \left [1-mx\right ]^+ \leq \tilde{y}_{i,k} < \left [1-(m-1) x-\Delta\right ]^+ \right \}.
\end{align*}
As $\tilde{\mathbf{x}}$ is fixed, the optimal solution would yield $\tilde{\mathbf{y}}$ {to be as small as} possible, while preserving feasibility.
{Thus,} $\forall (i,k) \in \tilde{\Pi}^{(m,j)}_{\Delta}, \tilde{y}_{i,k} = \left [1-m x - \Delta \right ]^+$ , and $\forall (i,k) \in \tilde{\Pi}^{(m,j)}_{\Delta,\dagger}, \tilde{y}_{i,k} = \left [1-m x \right ]^+$. 
By noticing that the sets $\tilde{\Pi}^{(m,j)}_{\Delta}$ and $\tilde{\Pi}^{(m,j)}_{\Delta,\dagger}$ are disjoint 
$\forall m \in [1:N]$ and 
$\forall j \in \left[0:\frac{N(N-1)}{2}\right]$ {and contain} all elements of $\tilde{\mathbf{y}}$, we can rewrite \eqref{eq11} as
{
\begin{align}
\label{f_tilde}
\sum\limits_{k=1}^{K} p_k & \sum\limits_{i=1}^N \tilde{y}_{i,k} = \!\!\!\!\!  \sum\limits_{j=0}^{\frac{N(N-1)}{2}} \!\!\!  B \left( p,j,N\right) \sum\limits_{m=1}^N  \left[ \sum\limits_{w \in \tilde{\Pi}^{(m,j)}_{\Delta}} \!\!\!\!\! \tilde{y}_{w} + \!\!\! \!\!\! \sum\limits_{w \in \tilde{\Pi}^{(m,j)}_{\Delta,\dagger}} \!\!\!\!\! \tilde{y}_{w} \right].
\end{align}}

Let
$Q_{N}^{m,j}\left(a,b\right) = a {N-1 \choose b - 1} {\frac{N(N-1)}{2}-N+1 \choose j - m + 1}$.
Then by means of counting techniques, one can see that the number of elements of $\tilde{\mathbf{y}}$ that belong to a constraint of the type $1-mx-\Delta$ is 
$\left |\tilde{\Pi}^{(m,j)}_{\Delta} \right | = Q_{N}^{m,j}\left( m,m \right)$,
while the number of elements of $\tilde{\mathbf{y}}$ that belong to a constraint of the type $1-mx$ is
$\left |\tilde{\Pi}^{(m,j)}_{\Delta,\dagger} \right | = Q_{N}^{m,j}\left(N-m,m \right) $.
With this we can rewrite \eqref{f_tilde} as in \eqref{f_tildeNew} at the top of the next page, where $g_1 \in \mathbb{Z}^+$, respectively $g_2 \in \mathbb{Z}^+$ 
which ensures that $\left [1-\ell x-\Delta \right ]^+=1-\ell x-\Delta, \forall \ell \in [1:g_1]$, respectively $\left [1-\ell x \right ]^+=1-\ell x, \forall \ell \in [1:g_2]$. 
\begin{figure*}
\begin{align}
\sum\limits_{k=1}^{K} p_k \sum\limits_{i=1}^N \tilde{y}_{i,k} &= \!\!\!  \sum\limits_{j=0}^{\frac{N(N-1)}{2}} \!\!\! B\left( p,j,N\right)  \left [\sum_{m=1}^N \left [ 1-mx-\Delta \right ]^+ Q_{N}^{m,j}\left(m,m \right) +\sum_{m=1}^N \left [ 1-mx \right ]^+ Q_{N}^{m,j}\left(N-m,m \right)  \right ] \nonumber
\\& = \!\!\! \sum\limits_{j=0}^{\frac{N(N-1)}{2}} \!\!\!\! B \left( p,j,N\right)  \left [\sum_{\ell=1}^{g_1} \left ( 1\!-\!\ell x\!-\!\Delta \right ) Q_{N}^{\ell,j}\left(\ell,\ell \right) \!+\!\sum_{\ell=1}^{g_2}  (1-\ell x ) Q_{N}^{\ell,j}\left(N-\ell,\ell \right)  \right ],
\label{f_tildeNew}
\\
\sum\limits_{k=1}^{K} p_k \sum\limits_{i=1}^N \hat{y}_{i,k} &\!= \!\!\! \sum\limits_{j=0}^{\frac{N(N-1)}{2}} \!\!\! B \left( p,j,N\right) \sum_{m=1}^N  \left [1\!-\!mx\!-\!\frac{m}{N} \Delta \right ]^+ \!\!\! Q_{N}^{m,j}\left(N,m \right) \!= \!\!\!\!\!  \sum\limits_{j=0}^{\frac{N(N-1)}{2}} \!\!\! B \left( p,j,N\right) \sum_{\ell=1}^{g_3} \left (1\!-\!\ell x\!-\!\frac{\ell}{N} \Delta \right ) Q_{N}^{\ell,j}\left(N,\ell \right).
\label{f_hatNew}
\end{align}
\hrulefill
\vspace{-5mm}
\end{figure*}
\\
\textbf{Evaluation of ${f}^{\text{Opt}}(\hat{\mathbf{x}},\hat{\mathbf{y}})$:}
We define 
 \begin{align*}
& \hat{\Pi}^{(m,j)}_{\Delta} = \left \{(i,k)| \ p_k = B \left( p,j,N\right),  \right.\\ 
 & \left. \left [1 - m x - \frac{m}{N} \Delta \right ]^+  \! \leq \! \hat{y}_{i,k} \!< \!\left [1 - (m-1)  x -  \frac{m-1}{N} \Delta \right ]^+  \right \}.
 \end{align*}
Thus, similarly to the case of $\left( \tilde{\mathbf{x}},\tilde{\mathbf{y}}\right )$, the optimal solution would yield $\forall (i,k) \in \hat{\Pi}^{(m,j)}_{\Delta}, \hat{y}_{i,k} = \left [1-m x - \frac{m}{N} \Delta \right ]^+$ and we can rewrite \eqref{eq11} as
{
\begin{align}
 \label{f_hat}
 \sum\limits_{k=1}^{K} p_k \sum\limits_{i=1}^N \hat{y}_{i,k} &= \sum\limits_{j=0}^{\frac{N(N-1)}{2}} B \left( p,j,N\right) 
{\cdot}\sum\limits_{m=1}^N \sum\limits_{w \in \hat{\Pi}^{(m,j)}_{\Delta}} \hat{y}_{w}.
\end{align}}
Similar to $\left( \tilde{\mathbf{x}},\tilde{\mathbf{y}}\right )$, one can see that the number of elements of $\hat{\mathbf{y}}$ that belong to a constraint of the type $1-mx-\frac{m}{N} \Delta$ is 
$\left |\hat{\Pi}^{(m,j)}_{\Delta} \right | = Q_{N}^{m,j}\left(N,m \right)$.
With this we have that \eqref{f_hat} can be rewritten as \eqref{f_hatNew} at the top of the next page, where $g_3 \in \mathbb{Z}^+$ ensures that $\left [1-\ell x - \frac{\ell}{N} \Delta \right ]^+ =1-\ell x - \frac{\ell}{N} \Delta, \forall \ell \in [1:g_3] $.

Depending on {$x$ and $\Delta$}, one can distinguish $4$ cases that might occur, which are next analyzed.
For each case, we identify the values of $g_{[1:3]}$ and show that ${f}^{\text{Opt}}(\tilde{\mathbf{x}},\tilde{\mathbf{y}}) - {f}^{\text{Opt}}(\hat{\mathbf{x}},\hat{\mathbf{y}}) \geq 0$, thus {proving} optimality of the pair $\left( \hat{\mathbf{x}},\hat{\mathbf{y}} \right)$. 
\\
\textbf{Case 1: $1-(n+1)x>0, n \in [1:N]$;} 
here, we have $g_1=g_3=n$ and $g_2=n+1$ in \eqref{f_tildeNew} and \eqref{f_hatNew} at the top of the next page. In that case we get ${{f}^{\text{Opt}}(\tilde{\mathbf{x}},\tilde{\mathbf{y}}) - {f}^{\text{Opt}}(\hat{\mathbf{x}},\hat{\mathbf{y}}) }\geq 0$.
\\
\textbf{Case 2: $1-(n+1)x\leq 0$ and $1-nx - \Delta >0, n \in [1:N]$;} 
here, we have $g_1 \! = \! g_2 \! = \! g_3 \! = \! n$ in \eqref{f_tildeNew} and \eqref{f_hatNew} at the top of the next page. In that case we get ${{f}^{\text{Opt}}(\tilde{\mathbf{x}},\tilde{\mathbf{y}}) - {f}^{\text{Opt}}(\hat{\mathbf{x}},\hat{\mathbf{y}}) } = 0$.
 \\
\textbf{Case 3: $1-nx - \Delta \leq 0$ and $1-nx - \frac{n}{N}\Delta >0, n \in [1:N]$;} here, we have $g_1 \!\! = \!\! n-1$ and $g_2 \!\! = \!\! g_3 \!\! = \!\! n$ in \eqref{f_tildeNew} and \eqref{f_hatNew} at the top of the next page. In that case we get ${{f}^{\text{Opt}}(\tilde{\mathbf{x}},\tilde{\mathbf{y}}) - {f}^{\text{Opt}}(\hat{\mathbf{x}},\hat{\mathbf{y}}) } \geq 0$.
\\ \textbf{Case 4: $1-nx - \frac{n}{N}\Delta \leq 0$ and $1- nx >0, n \in [1:N]$;} here, we have $g_1=g_3=n-1$ and $g_2=n$ in \eqref{f_tildeNew} and \eqref{f_hatNew} at the top of the page. In that case we get  ${{f}^{\text{Opt}}(\tilde{\mathbf{x}},\tilde{\mathbf{y}}) - {f}^{\text{Opt}}(\hat{\mathbf{x}},\hat{\mathbf{y}}) } \geq 0$.
\begin{remark}
{The proof above generalizes to the case when $\tilde{\mathbf{x}}$} has general components. The idea is {to order $\tilde{\mathbf{x}}$} in ascending order and to rewrite it as {$\tilde{\mathbf{x}}= x_m+ \left[ 0, \ \Delta_2, \ \hdots, \ \Delta_N \right]^T$, }
 where $x_m = \min_i \tilde{\mathbf{x}}_i$ and $\Delta_i, \ i \in [1:N]$ is the difference between the $i$-th component of the ordered $\tilde{\mathbf{x}}$ and $x_m$. Then, the above method is applied $N-1$ times as follows. At step $k \in [1:N-1]$, $\Delta_{k+1}$ is equally shared among the $N$ components of $\tilde{\mathbf{x}}$; this, as proved above, brings to a reduction of the objective function. At the end of the $N-1$ steps the optimal $\hat{\mathbf{x}}$ is of the form $\hat{\mathbf{x}}= \mathbf{1}_N \hat{x}$. 
\end{remark} 

According to the structure of $\hat{\mathbf{x}}$ in Step 2, the optimal {$\mathbf{\hat{y}}$} has components of the form {stated} in Lemma \ref{lemma:optSol} - item 2).

{One can} see that with $f^{\text{Opt}}\left( \hat{\mathbf{x}}, \hat{\mathbf{y}} \right)$ in \eqref{opt_problem_original_no_relay}, we have
\begin{align*}
\sum\limits_{k=1}^{K} p_k \sum\limits_{i=1}^{N} \hat{y}_{i,k} = \sum\limits_{i=1}^{N} \hat{y}_i {N-1 \choose i-1} N p^{i-1} (1-p)^{N-i},
\end{align*}
which follows by noting that the probability of having $i$ people meeting is ${N-1 \choose i-1} p^{i-1} (1-p)^{N-i}$ and that this event happens once for every user. 
This completes the proof of Lemma \ref{lemma:optSol}.

{Using Lemma \ref{lemma:optSol}}, one can prove the equivalence of the LPs in \eqref{opt_problem_original_no_relay} and \eqref{opt_problem_general_N_no_relay_simplified} 
using similar arguments as in Appendix \ref{app:proofOfLPreduction}.}
This concludes the proof.

\section{Proof of Theorem \ref{theorem_set_cover}}
\label{app:theorem_set_cover}
Let $\bar{f}^{\text{Opt}}$ be the optimal solution for the LP in \eqref{opt_problem_original_no_relay}. 
Then, the LP in \eqref{opt_problem_original_no_relay} can be equivalently written as
\begin{equation}
\nonumber
\begin{aligned}
\bar{f}^{\text{Opt}} = & \min\limits_{\left(\mathbf{x},\mathbf{y} \right) \in \mathcal{F}} \sum_{k =1}^K p_k \sum_{i=1}^N y_{i,k} + \mathbf{1}_N^T\mathbf{x},
\end{aligned}
\end{equation}
where 
$ \mathcal{F} = \{ \left(\mathbf{x}, \mathbf{y} \right) | \mathbf{x} \geq \mathbf{0}_N, \: \: \mathbf{y} \geq \mathbf{0}_{N \times K},
 \mathbf{1}_N - \mathbf{A}^{(k)} \mathbf{x} \leq \mathbf{y}_k, \forall \ k\in[1:K]\}$.
The next series of inequalities hold
\begin{equation}
 \nonumber
\begin{aligned}
 \bar{f}^{\text{Opt}} & = \min\limits_{\left(\mathbf{x},\mathbf{y} \right) \in \mathcal{F}} \sum_{k =1}^K p_k \sum_{i=1}^N y_{i,k} + \mathbf{1}_N^T\mathbf{x} \\
 & = \min\limits_{\left(\mathbf{x},\mathbf{y} \right) \in \mathcal{F}} \sum_{k =1}^K p_k \sum_{i=1}^N y_{i,k} +  \sum_{k =1}^K p_k \mathbf{1}_N^T\mathbf{x} \\
 &\stackrel{(a)}{\geq} \sum_{k =1}^K p_k  \min\limits_{\left(\mathbf{x},\mathbf{y} \right) \in \mathcal{F}} \left( \sum_{i=1}^N y_{i,k} +   \mathbf{1}_N^T\mathbf{x} \right) =  \sum_{k =1}^K p_k f(k),\\
\end{aligned}
\end{equation}
where $(a)$ is due to Jensen's inequality. Thus, 
\begin{equation}
\label{lb_set}
\begin{aligned}
f(k) = &\min\limits_{\mathbf{x},\mathbf{y}} & & \sum_{i=1}^N y_{i,k} + \mathbf{1}_N^T\mathbf{x}  \\
&\text{subject to } & & \mathbf{x} \geq \mathbf{0}_N, \: \: \mathbf{y} \geq \mathbf{0}_{N \times K},  \\
&   & & \mathbf{1}_N - \mathbf{A}^{(v)} \mathbf{x} \leq \mathbf{y}_v, \forall \ v\in[1:K].
\end{aligned}
\end{equation}
Consider the set of constraints in \eqref{lb_set} $\forall v \in [1:K]$ such that $v \neq k$, which can be written as
{$ \mathbf{A}^{(v)}  \mathbf{x} \geq \mathbf{1}_N - \mathbf{y}_v  \geq -\infty$},
since $\mathbf{y}_v$ {does} not directly affect the objective function. This makes these constraints trivial, i.e., \eqref{lb_set} becomes
\begin{equation}
\label{lb_set_simplified}
\begin{aligned}
f(k) = &\min\limits_{\mathbf{x},\mathbf{y}_k} & & \mathbf{1}^T_N \mathbf{y}_k + \mathbf{1}_N^T\mathbf{x} \\
&\text{subject to } & & \mathbf{x} \geq \mathbf{0}_N, \: \: \mathbf{y}_k \geq \mathbf{0}_{N},  
\\
&   & & 
\mathbf{y}_k + \mathbf{A}^{(k)} \mathbf{x} \geq \mathbf{1}_N. 
\end{aligned}
\end{equation}
For the problem in \eqref{lb_set_simplified}, we prove that, for any $k \in [1:K]$, an optimal solution of the form $\mathbf{y}_k = \mathbf{0}_N$ always exists. 
Assume this is not true, i.e., $\exists \left(\mathbf{x}^*, \mathbf{y}_k^* \right) \ \text{such \ that} \ {y}_{i,k} > 0, \ \text{for} \  i \in \mathcal{S}, \ \text{where} \ \mathcal{S} \subseteq [1:N]$.
Then consider the point $\left(\hat{\mathbf{x}},\hat{\mathbf{y}}_k \right)$. Note that $\mathbf{A}^{(k)}$ has ones on the diagonal. 
Then, by letting $\hat{\mathbf{y}}_k = \mathbf{0}_N$ and $\hat{\mathbf{x}} = \mathbf{x}^* + \mathbf{y}_k^*$ we get another feasible point with the same objective function. 
Thus, the problem in \eqref{lb_set_simplified} becomes
\begin{equation}
\label{lb_set_cover}
\begin{aligned}
f(k) = &\min\limits_{\mathbf{x}} & & \mathbf{1}_N^T\mathbf{x} \\
&\text{subject to } & & \mathbf{x} \geq \mathbf{0}_N,  \ \mathbf{A}^{(k)} \mathbf{x} \geq \mathbf{1}_N. 
\end{aligned}
\end{equation}
The problem in \eqref{lb_set_cover} is the LP relaxation of the SC problem on a bipartite graph with adjacency matrix $\mathbf{A}^{(k)}$.
This concludes the proof.

\section{Proof of Theorem \ref{th:coveringSymm}}
\label{app:Covering}
We start by proving that $\mathbf{x}^{\text{PSC}} = x \mathbf{1}_N$. 
Assume that $\mathbf{x}^{\prime}$ is the optimal solution, with $k= \min_{i \in [1:N]} \left \{\mathbf{x}^{\prime}_i \right \}$; without loss of generality, $\mathbf{x}^{\prime} = k \mathbf{1}_N + [\Delta_{[1:N-1]}, \: 0 ]^T$, where $\Delta_i \geq 0 \ \forall i=[1:N-1]$.
{With this, we get that}
${f^{\text{PSC}}(\mathbf{x}^{\prime})} = N k+ \sum\limits_{i=1}^{N-1} \Delta_i$,
and ${\mathbf{x}}^{\prime}$ satisfies the constraints of the LP in \eqref{covering_problem} that are
\begin{align}
\label{eq:generalcons}
&k + (N-1) p k + \Delta_i + p\sum\limits_{j=1,j\neq i}^{N-1} \Delta_j \geq 1, \ {\forall i \in [1:N-1]} \nonumber \\
&k + (N-1)pk + p \sum\limits_{j=1}^{N-1} \Delta_j \geq 1.
\end{align}
It is clear that the first set of constraints is always redundant, as the second one is tighter. Now consider $\mathbf{x}^{\text{PSC}} = x \mathbf{1}_N$, with {$x = k+\frac{1}{N} \sum\limits_{j=1}^{N-1} \Delta_j$};
it is not difficult to see that {$f^{\text{PSC}}(\mathbf{x}^{\text{PSC}}) = f^{\text{PSC}}(\mathbf{x}^{\prime})$}. 
To complete the proof that $\mathbf{x}^{\text{PSC}} = x \mathbf{1}_N$ we need to show that such a point is feasible. The constraints of the LP in \eqref{covering_problem}  when evaluated at $\mathbf{x}^{\text{PSC}}$ become
\begin{equation*}
k + (N-1)pk + \frac{1}{N} \left( 1+\left( N-1\right)p\right) \sum\limits_{j=1}^{N-1} \Delta_j \geq 1,
\end{equation*}
thus $\mathbf{x}^{\text{PSC}}$ is a feasible solution {as this constraint is always satisfied if the second constraint in \eqref{eq:generalcons} holds since $\frac{1}{N} \left( 1+\left( N-1\right)p\right) \geq p$.}
By enforcing this solution into the LP in \eqref{covering_problem} 
we get $f^{\text{PSC}}\left( {\mathbf{x}}^{\text{PSC}}\right) = N x$ and a constraint of the form $x {\geq} \frac{1}{1 + (N-1)p} = \frac{1}{\mathbb{E}(C)}$ which implies that the optimal value is $\mathbf{x}^{\text{PSC}} = \frac{1}{\mathbb{E}(C)} \mathbf{1}_N$. This completes the proof.

\section{Proof of Theorem \ref{theorem_lb}}
\label{app:theorem_lb} 
 Let $\left({\mathbf{x}}^{\text{Opt}}, {\mathbf{y}}^{\text{Opt}} \right )$ and ${\mathbf{x}}^{\text{PSC}}$ be the optimal points for the LPs in \eqref{opt_problem_original_no_relay} and \eqref{covering_problem}, respectively.
 If ${\mathbf{x}}^{\text{Opt}}$ is a feasible point in the LP in \eqref{covering_problem}, then by definition
 \begin{align}
  \nonumber
  \bar{f}^{\text{PSC}} = f^{\text{PSC}}(\mathbf{x}^{\text{PSC}}) \leq f^{\text{PSC}}(\mathbf{x}^{\text{Opt}}) \leq f^{\text{Opt}}(\mathbf{x}^{\text{Opt}},\mathbf{y}^{\text{Opt}}) = {\bar{f}}^{\text{Opt}},
 \end{align}
 \noindent where the second inequality holds by simply observing the objective functions of both problems.
 
Consider now the case where $\mathbf{x}^{\text{Opt}}$ is not a feasible point in the LP in \eqref{covering_problem}. Since it is feasible in \eqref{opt_problem_original_no_relay}, it satisfies
  \begin{equation}
   \label{consts}
   \mathbf{A}^{(k)}\mathbf{x}^{\text{Opt}} + \mathbf{y}_k^{\text{Opt}} \geq \mathbf{1}_N, \: \forall k \in [1:K].
  \end{equation}
Weighting \eqref{consts} by $p_k$ and taking the sum, we get
  \begin{equation}
\label{eq:greatOne}
   \sum\limits_{k=1}^{K} p_k \left(\mathbf{A}^{(k)}\mathbf{x}^{\text{Opt}} + \mathbf{y}_k^{\text{Opt}}\right ) = \mathbf{P}\mathbf{x}^{\text{Opt}} + \mathbf{\tilde{y}}^{\text{Opt}} \geq \mathbf{1}_N
  \end{equation}
  where the equality holds by noticing that $\sum\limits_{k=1}^{K} p_k \mathbf{A}^{(k)} = \mathbf{P}$ and by letting $ \mathbf{\tilde{y}}^{\text{Opt}} = \sum\limits_{k=1}^{K} p_k \mathbf{y}_k^{\text{Opt}} \geq \mathbf{0}_N$.
  
Consider now the point $\mathbf{\hat{x}}^{\text{PSC}} = \mathbf{x}^{\text{Opt}} + \mathbf{\tilde{y}}^{\text{Opt}}$. Then, this point is feasible in \eqref{covering_problem} since
 $  \mathbf{P}\mathbf{\hat{x}}^{\text{PSC}} = \mathbf{P}\mathbf{x}^{\text{Opt}} + \mathbf{P} \mathbf{\tilde{y}}^{\text{Opt}} 
   \geq \mathbf{P}\mathbf{x}^{\text{Opt}} + \mathbf{\tilde{y}}^{\text{Opt}} \geq \mathbf{1}_N$,
where the first inequality holds because the diagonal entries of $\mathbf{P}$ are all equal to 1, and all the other entries are non-negative, while the second inequality follows from \eqref{eq:greatOne}. Thus, we get
   \begin{align}
  \nonumber
   \bar{f}^{\text{PSC}} \leq f^{\text{PSC}}(\mathbf{\hat{x}}^{\text{PSC}}) = f^{\text{Opt}}(\mathbf{x}^{\text{Opt}},\mathbf{y}^{\text{Opt}}) = \bar{f}^{\text{Opt}}.
 \end{align}
This completes the proof.

%
%
%

\ifCLASSOPTIONcaptionsoff
  \newpage
\fi



\bibliographystyle{IEEEtran}
\bibliography{JournalBib}

\begin{thebibliography}{10}
\providecommand{\url}[1]{#1}
\csname url@samestyle\endcsname
\providecommand{\newblock}{\relax}
\providecommand{\bibinfo}[2]{#2}
\providecommand{\BIBentrySTDinterwordspacing}{\spaceskip=0pt\relax}
\providecommand{\BIBentryALTinterwordstretchfactor}{4}
\providecommand{\BIBentryALTinterwordspacing}{\spaceskip=\fontdimen2\font plus
\BIBentryALTinterwordstretchfactor\fontdimen3\font minus
  \fontdimen4\font\relax}
\providecommand{\BIBforeignlanguage}[2]{{%
\expandafter\ifx\csname l@#1\endcsname\relax
\typeout{** WARNING: IEEEtran.bst: No hyphenation pattern has been}%
\typeout{** loaded for the language `#1'. Using the pattern for}%
\typeout{** the default language instead.}%
\else
\language=\csname l@#1\endcsname
\fi
#2}}
\providecommand{\BIBdecl}{\relax}
\BIBdecl

\bibitem{maddah2014fundamental}
M.~A. Maddah-Ali and U.~Niesen, ``Fundamental limits of caching,'' \emph{IEEE
  Trans. Inf. Theory}, vol.~60, no.~5, pp. 2856--2867, 2014.

\bibitem{hachem2014multi}
J.~Hachem, N.~Karamchandani, and S.~Diggavi, ``Multi-level coded caching,'' in
  \emph{2014 IEEE International Symposium on Information Theory (ISIT)}, pp.
  56--60.

\bibitem{ji2016wireless}
M.~Ji, G.~Caire, and A.~F. Molisch, ``Wireless device-to-device caching
  networks: Basic principles and system performance,'' \emph{IEEE J. Sel. Areas
  Commun.}, vol.~34, no.~1, pp. 176--189, 2016.

\bibitem{ananthanarayanan2007combine}
G.~Ananthanarayanan, V.~N. Padmanabhan, L.~Ravindranath, and C.~A. Thekkath,
  ``Combine: leveraging the power of wireless peers through collaborative
  downloading,'' in \emph{Proceedings of the 5th international conference on
  Mobile systems, applications and services}.\hskip 1em plus 0.5em minus
  0.4em\relax ACM, 2007, pp. 286--298.

\bibitem{keller2012microcast}
L.~Keller, A.~Le, B.~Cici, H.~Seferoglu, C.~Fragouli, and A.~Markopoulou,
  ``Microcast: cooperative video streaming on smartphones,'' in
  \emph{Proceedings of the 10th international conference on Mobile systems,
  applications, and services}.\hskip 1em plus 0.5em minus 0.4em\relax ACM,
  2012, pp. 57--70.

\bibitem{do2011massive}
N.~M. Do, C.-H. Hsu, J.~P. Singh, and N.~Venkatasubramanian, ``Massive live
  video distribution using hybrid cellular and ad hoc networks,'' in \emph{2011
  IEEE International Symposium on a World of Wireless, Mobile and Multimedia
  Networks (WoWMoM)}, pp. 1--9.

\bibitem{chun2004selfish}
B.-G. Chun, K.~Chaudhuri, H.~Wee, M.~Barreno, C.~H. Papadimitriou, and
  J.~Kubiatowicz, ``Selfish caching in distributed systems: a game-theoretic
  analysis,'' in \emph{Proceedings of the twenty-third annual ACM symposium on
  Principles of distributed computing}, 2004, pp. 21--30.

\bibitem{goemans2006market}
M.~X. Goemans, L.~Li, V.~S. Mirrokni, and M.~Thottan, ``Market sharing games
  applied to content distribution in ad hoc networks,'' \emph{IEEE J. Sel.
  Areas Commun.}, vol.~24, no.~5, pp. 1020--1033, 2006.

\bibitem{taghizadeh2013distributed}
M.~Taghizadeh, K.~Micinski, C.~Ofria, E.~Torng, and S.~Biswas, ``Distributed
  cooperative caching in social wireless networks,'' \emph{IEEE Trans. Mob.
  Comput.}, vol.~12, no.~6, pp. 1037--1053, 2013.

\bibitem{reich2009age}
J.~Reich and A.~Chaintreau, ``The age of impatience: optimal replication
  schemes for opportunistic networks,'' in \emph{Proceedings of the 5th
  international conference on Emerging networking experiments and
  technologies}.\hskip 1em plus 0.5em minus 0.4em\relax ACM, 2009, pp. 85--96.

\bibitem{ioannidis2010distributed}
S.~Ioannidis, L.~Massoulie, and A.~Chaintreau, ``Distributed caching over
  heterogeneous mobile networks,'' in \emph{ACM SIGMETRICS Performance
  Evaluation Review}, vol.~38, no.~1, 2010, pp. 311--322.

\bibitem{tajbakhsh2015delay}
S.~E. Tajbakhsh and P.~Sadeghi, ``Delay tolerant information dissemination via
  coded cooperative data exchange,'' \emph{Journal of Communications and
  Networks}, vol.~17, no.~2, pp. 133--144, 2015.

\bibitem{rebecchi2015data}
F.~Rebecchi, M.~Dias~de Amorim, V.~Conan, A.~Passarella, R.~Bruno, and
  M.~Conti, ``Data offloading techniques in cellular networks: a survey,''
  \emph{IEEE Communications Surveys \& Tutorials}, vol.~17, no.~2, pp.
  580--603, 2015.

\bibitem{han2012mobile}
B.~Han, P.~Hui, V.~A. Kumar, M.~V. Marathe, J.~Shao, and A.~Srinivasan,
  ``Mobile data offloading through opportunistic communications and social
  participation,'' \emph{IEEE Trans. Mob. Comput.}, vol.~11, no.~5, pp.
  821--834, 2012.

\bibitem{barbera2014data}
M.~V. Barbera, A.~C. Viana, M.~D. De~Amorim, and J.~Stefa, ``Data offloading in
  social mobile networks through vip delegation,'' \emph{Ad Hoc Networks},
  vol.~19, pp. 92--110, 2014.

\bibitem{Medard2006}
S.~Deb, M.~Medard, and C.~Choute, ``Algebraic gossip: a network coding approach
  to optimal multiple rumor mongering,'' \emph{IEEE Trans. Inf. Theory},
  vol.~52, no.~6, pp. 2486--2507, 2006.

\bibitem{pentland2009inferring}
A.~Pentland, N.~Eagle, and D.~Lazer, ``Inferring social network structure using
  mobile phone data,'' \emph{Proceedings of the National Academy of Sciences
  (PNAS)}, vol. 106, no.~36, pp. 15\,274--15\,278, 2009.

\bibitem{kosta2010small}
S.~Kosta, A.~Mei, and J.~Stefa, ``Small world in motion (swim): Modeling
  communities in ad-hoc mobile networking,'' in \emph{2010 7th Annual IEEE
  Communications Society Conference on Sensor Mesh and Ad Hoc Communications
  and Networks (SECON)}, pp. 1--9.

\bibitem{hui2005pocket}
P.~Hui, A.~Chaintreau, J.~Scott, R.~Gass, J.~Crowcroft, and C.~Diot, ``Pocket
  switched networks and human mobility in conference environments,'' in
  \emph{Proceedings of the 2005 ACM SIGCOMM workshop on Delay-tolerant
  networking}, pp. 244--251.

\bibitem{chaintreau2005pocket}
A.~Chaintreau, P.~Hui, J.~Crowcroft, C.~Diot, R.~Gass, and J.~Scott, ``Pocket
  switched networks: Real-world mobility and its consequences for opportunistic
  forwarding,'' Technical Report UCAM-CL-TR-617, University of Cambridge,
  Computer Laboratory, Tech. Rep., 2005.

\bibitem{karmooseISIT2016}
M.~Karmoose, M.~Cardone, and C.~Fragouli, ``Simplifying wireless social
  caching,'' in \emph{2016 IEEE International Symposium on Information Theory
  (ISIT)}.

\bibitem{ekman2008working}
F.~Ekman, A.~Ker{\"a}nen, J.~Karvo, and J.~Ott, ``Working day movement model,''
  in \emph{Proceedings of the 1st ACM SIGMOBILE workshop on Mobility models},
  2008, pp. 33--40.

\bibitem{karamshuk2011human}
D.~Karamshuk, C.~Boldrini, M.~Conti, and A.~Passarella, ``Human mobility models
  for opportunistic networks,'' \emph{IEEE Commun. Mag.}, vol.~49, no.~12, pp.
  157--165, 2011.

\bibitem{gonzalez2008understanding}
M.~C. Gonzalez, C.~A. Hidalgo, and A.-L. Barabasi, ``Understanding individual
  human mobility patterns,'' \emph{Nature}, vol. 453, no. 7196, pp. 779--782,
  2008.

\bibitem{milgram1967small}
S.~Milgram, ``The small world problem,'' \emph{Psychology today}, vol.~2,
  no.~1, pp. 60--67, 1967.

\bibitem{beraldi2002probabilistic}
P.~Beraldi and A.~Ruszczynski, ``The probabilistic set-covering problem,''
  \emph{Operations Research}, vol.~50, no.~6, pp. 956--967, 2002.

\bibitem{chaintreau2007impact}
A.~Chaintreau, P.~Hui, J.~Crowcroft, C.~Diot, R.~Gass, and J.~Scott, ``Impact
  of human mobility on opportunistic forwarding algorithms,'' \emph{IEEE Trans.
  Mob. Comput.}, vol.~6, no.~6, pp. 606--620, 2007.

\end{thebibliography}

\begin{IEEEbiography}[{\includegraphics[width=1in,height=1.25in,clip,keepaspectratio]{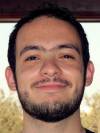}}]{Mohammed Karmoose} is a Ph.D student in the Electrical Engineering department at UCLA. He received the BS and MS degrees in electrical engineering from the Faculty of Engineering in Alexandria University in Egypt in 2009 and 2013 respectively. 
He was a part of the CRN research group in E-JUST in Egypt as a Graduate Research Assistant from 2011 to 2014.
He received the Annual Tribute Ceremony award for top-ranked students in Alexandria University in the years 2005 to 2009.
He received the Electrical Engineering Department Fellowship from UCLA for his first year of Ph.D in 2014/2015.
His research interests are distributed detection, cooperative caching and wireless communications.

\end{IEEEbiography}

\begin{IEEEbiography}[{\includegraphics[width=1in,height=1.25in,clip,keepaspectratio]{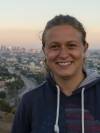}}]{Martina Cardone}
is currently a post-doctoral research fellow in the Electrical Engineering department at UCLA. She received the B.S. (Telecommunications Engineering) and the M.S. (Telecommunications Engineering) degrees summa cum laude from the Politecnico di Torino, Italy, in 2009 and 2011, respectively and the M.S. in Telecommunications Engineering from Télécom ParisTech, Paris, France, in 2011, as part of a double degree program. In 2015, she received the Ph.D. in Electronics and Communications from Télécom ParisTech (with work done at Eurecom in Sophia Antipolis, France). She received the second prize in the Outstanding Ph.D. award, Télécom ParisTech, Paris, France and the Qualcomm Innovation Fellowship in 2014. Her research interests are in network information theory, content-type coding, cooperative caching and wireless network secrecy.
\end{IEEEbiography}

\begin{IEEEbiography}[{\includegraphics[width=1in,height=1.25in,clip,keepaspectratio]{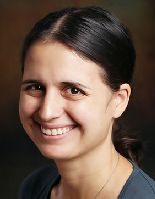}}]{Christina Fragouli} is a Professor in the Electrical Engineering department at UCLA. She received the B.S. degree in Electrical Engineering from the National Technical University of Athens, Athens, Greece, in 1996, and the M.Sc. and Ph.D. degrees in Electrical Engineering from the University of California, Los Angeles, in 1998 and 2000, respectively. She has worked at the Information Sciences Center, AT\&T Labs, Florham Park New Jersey, and the National University of Athens. She also visited Bell Laboratories, Murray Hill, NJ, and DIMACS, Rutgers University.
Between 2006--2015 she was faculty  in the School of Computer and Communication Sciences, EPFL, Switzerland. She is an IEEE fellow, has served as an Information Theory Society Distinguished Lecturer, and as an Associate Editor for IEEE Communications Letters, Elsevier Journal on Computer Communication, IEEE Transactions on Communications, IEEE Transactions on Information Theory, and IEEE Transactions on Mobile Communications. Her research interests are in network coding, wireless communications, algorithms for networking and network security.
\end{IEEEbiography}





\end{document}